\documentclass[a4paper,twocolumn,showpacs,superscriptaddress,floatfix]{iopart}

\usepackage{iopams}
\expandafter\let\csname equation*\endcsname\relax
\expandafter\let\csname endequation*\endcsname\relax
\usepackage{amsmath}

\usepackage{soul}

\usepackage{graphicx}
\usepackage{epsf}
\usepackage{epsfig}
\usepackage[usenames]{color}
\usepackage[dvipsnames]{xcolor}
\usepackage{times}

\usepackage{amssymb}
\usepackage{slashed}
\usepackage{hyperref}
\usepackage{comment}

\newcommand{\be}{\begin{equation}}
\newcommand{\ee}{\end{equation}}
\newcommand{\bea}{\begin{eqnarray}}
\newcommand{\eea}{\end{eqnarray}}
\newcommand{\bal}{\begin{align}}
\newcommand{\eal}{\end{align}}

\newcommand{\Hor}{{\mathcal H}}

\font\tenscr=rsfs10 scaled1100
\font\sevenscr=rsfs7 
\font\fivescr=rsfs5 
\skewchar\tenscr='177
\skewchar\sevenscr='177
\skewchar\fivescr='177
\newfam\scrfam
\textfont\scrfam=\tenscr
\scriptfont\scrfam=\sevenscr
\scriptscriptfont\scrfam=\fivescr

\def\scri{{\fam\scrfam I}}

\usepackage[caption=false]{subfig}
\usepackage{url}

\usepackage{xcolor}
\usepackage{color}

\begin{document}

\title{Vishveshwara's waveform revisited: insights from a Keldysh quasinormal mode expansion}

\author{J. Besson$^{1,2,3}$ and J.L. Jaramillo $^1$}

\address{${^1}$ Institut de Math\'ematiques de Bourgogne UMR 5584,
  Universit\'e Bourgogne Europe, CNRS, F-21000 Dijon, France}

\address{${^2}$ Albert-Einstein-Institut, Max-Planck-Institut für Gravitationsphysik, Callinstraße 38, 30167 Hannover, Germany}

    \address{${^3}$Leibniz Universit\"at Hannover, 30167 Hannover, Germany}

\begin{abstract}
We revisit Vishveshwara's linear scattering on a Schwarzschild black hole to study the evolution of quasinormal mode activation along the full time-domain waveform. Specifically, by adopting a hyperboloidal approach we cast the scattering problem in a non-selfadjoint dynamics setting where a (Keldysh) resonant expansion in a bi-orthogonal system of quasinormal modes can be readily performed. The calculation reveals the neat correlation of the second peak in Vishveshwara's waveform to the fundamental mode and first overtones, closely following the ringdown waveform pattern of non-linear evolutions. The first peak is completely controlled, for even (Zerilli) perturbations, by the algebraically special mode and the nearby branch cut. This phenomenon is absent for odd (Regge-Wheeler) perturbations. These qualitative features are robust under change of initial data, extending to Reissner-Nordstr\"om and Kerr black holes but being absent when departing from asymptotic flatness to spacetime asymptotics without a branch cut. 
Underlying linear mechanisms
might shed light on the role of algebraic speciality in the transition to the ringdown in non-linear evolutions.

\end{abstract}

\section{Introduction: on Vishveshwara's black hole trail}
\label{s:intro}
Vishveshwara's 1970 pioneering work  
\cite{vishveshwara1970scattering}, on the (linear) scattering of wavepackets by a Schwarzschild's black hole, marked the beginning of a long-extended effort to model black hole (BH) dynamics and to understand its underlying mechanisms. His identification of a ``saturated pattern" in the scattered waves (now the celebrated Vishveshwara's waveform) from an incident wavepacket with width  of the order of (or smaller than) the BH lengthscale, led to the recognition of the BH imprint  in the scattered radiation, in particular the BH mass, thus opening the pathway to gravitational wave astrophysics. From a structural perspective,  the realisation of the key role of outgoing boundary conditions led him to the
understanding of ``radiation damping and resonance scattering"  (cf. Vishveshwara's insights in~\cite{Vishveshwara1968,vishveshwara1970stability,Vishveshwara_CurrentScience:1996jgz}) as the dissipation mechanism underlying BH stability, in particular
identifying for the first time the ``ringdown waves" that would give rise
to the notion of quasinormal modes (QNM). Indeed, shortly after Vishveshwara's time-domain calculation, Chandrasekhar and Detweiler completed this BH resonance picture with the spectral-domain computation of QNM complex frequencies~\cite{ChaDet75} (see~\cite{Berti:2009kk} for a detailed account). Since these pioneer works, a large and sound body of knowledge has been established, in particular addressing the full-nonlinear problem. Glossing over more than five decades of steady work and results, some key milestones in this research trail are: the binary black hole merger numerical simulations by 
Pretorius~\cite{Pre05}, the detection of gravitational waves including the
predicted QNMs by Advanced LIGO~\cite{Abbott:2016blz} and, recently, the claimed proofs of non-linear stability for (subextremal) Kerr BHs
\cite{hintz2026nonlinear,Szeftel:2026qzr}.

In spite of the definite success of this overall picture, important specific aspects (such as identifying the onset of the ringdown regime or the mechanism underlying the simplicity and universality of BH merger waveforms~\cite{Jaramillo:2023day}) remain to be elucidated.
In this note we get back to Vishveshwara's methodology in \cite{vishveshwara1970scattering}, revisiting his linear dynamics scattering approach but fully combining his
time-domain scheme with the spectral-domain analysis fostered   in~\cite{ChaDet75}. Specifically, adopting a hyperboloidal approach, we perform a complete (Keldysh) resonant 
expansion~\cite{besson_quasi-normal_2025}
 of linear Vishveshwara-like waveforms into QNMs and branch cut
 contribution. We report on the results, monitoring the activation of QNMs in time, with an emphasis on the gained insights into the ringdown onset and on the (closely related)  role of the algebraically special (AS) QNM in the (universal) transition from the prompt signal to the ringdown regime.

\section{Hyperboloidal slicing and non-selfadjoint dynamics}
\label{s:hyper}
From a technical perspective, the key shift with respect to~\cite{vishveshwara1970scattering} is the adoption of a hyperboloidal slicing scheme, that permits i) to write the dynamics explicitly as the flow of a non-selfadjoint infinitesimal time generator and, crucially, ii) to perform a full
QNM waveform decomposition by 
solving the non-selfadjoint spectral problem of such operator.

Closely following~\cite{besson_quasi-normal_2025}, we start by 
writing the wave equation for a scalar master equation
\bea
\left(\frac{\partial^2}{\partial t^2} - \frac{\partial^2}{\partial r_*^2} + V \right)\phi=0 \ ,
\label{wave_equation_tortoise}
    \eea
 where $V$ is the corresponding BH effective potential and $r_*$ is a tortoise coordinate. This Cauchy slice formulation is the starting point for Vishveshwara's analysis in Schwarzschild, where $V$ is the Regge-Wheeler or the Zerilli potential for an $(\ell, m)$ mode, subject to outgoing boundary conditions at the BH horizon $r_*\to-\infty$ and at (spatial) infinity $r_*\to\infty$. Rescaling the variables as
    $\bar t = t/\lambda,\bar x = r_*/\lambda\in]-\infty, \infty[, \hat V = \lambda^2 V$, for a convenient lengthscale $\lambda$, the hyperboloidal scheme~\cite{Zenginoglu:2011jz,Warnick:2013hba,Ansorg:2016ztf,MacZen25} is built by performing
    the change of coordinates 
\bea
\label{e:hyperboloidal_change}
\begin{cases}
  	\bar{t} = \tau - h(x)\\
  	\bar{x} = g(x)
\end{cases} \ ,
\label{compactified_hyperboloidal}
\eea
for an appropriate height function $h(x)$, such  that $\tau=\mathrm{const}$ slices are asymptotically hyperboloidal spatial hypersurfaces interpolating between the BH horizon and null infinity, and with $g(x)$ a (coordinate) compactification
to an interval $[a,b]$ along such hyperboloidal slices (points
$a$ and $b$ are added to include the BH horizon and null infinity in the domain; outgoing boundary conditions translate into regularity conditions at $a$ and $b$). Performing a first-order reduction in time $\psi = \partial_\tau \phi$, Eq. (\ref{wave_equation_tortoise}) can be written~\cite{Jaramillo:2020tuu,besson_quasi-normal_2025} in the following form
\begin{equation}
\label{e:wave_eq_1storder_u_tau}
\displaystyle\left\{
\begin{array}{l}
 \partial_\tau u  = i L  u \ , \\
 u(\tau,x)\vert_{\tau=0}=u_0(x) \ \ , \ \ 
 \end{array}
 \right. 
\end{equation}
with the propagating (scattered) field $u$ and its initial data $u_0$ given by 
\bea
        u(\tau,x) = \begin{pmatrix} \phi(\tau,x)\\ \psi(\tau, x) \end{pmatrix}  \ \ , \  \   u_0(x) = \begin{pmatrix}   \phi_0(x) = \phi(\tau, x)\vert_{\tau=0}\\   \psi_0(x) = \partial_\tau\phi(\tau, x)\vert_{\tau=0} \end{pmatrix}  \ ,
        \eea  
and where the infinitesimal time generator is the non-selfadjoint operator 
\bea
\label{e:L_L_1_L_2_main}
	 L = \frac{1}{i}
	\left(\begin{array}{c|c}
    0 & 1\\ \hline
    L_1 & L_2
	\end{array}\right) \ ,
\eea
with $L_2$ encoding the fluxes through the boundaries,
thus accounting for $L$ non-selfadjointness (cf. \cite{Jaramillo:2020tuu,besson_quasi-normal_2025} for the explicit expressions of the operators  $L_1$ and $L_2$ in terms of 
 $V$, $h$ and $g$).

\section{Keldysh QNM resonant expansion}
Solving Eq. (\ref{e:wave_eq_1storder_u_tau}) provides the time-domain waveform resulting from linear BH scattering. Our main contribution here stems however from the spectral-domain analysis, namely the QNM expansion of the waveform~\cite{besson_quasi-normal_2025}. We consider the spectral problem of $L$ and its transpose $L^t$
\bea
       \label{e:eigen_L-Lt_conclusions}
       L v_n = \omega_n v_n \ \ , \ \  L^t \alpha_n = \omega_n \alpha_n
       \ \ , \ \ v_n\in{\cal B}, \alpha_n\in{\cal B}^* \ ,
       \eea
with ${\cal B}$ an appropriate Banach space and ${\cal B}^*$ its dual (note that no scalar product is a priori introduced). Under the assumption, satisfied in our problem, that eigenvalues 
$\omega_n$ are simple, we can choose modes $v_n$ and comodes
$\alpha_n$ to form a bi-orthogonal system, namely satisfying 
 \bea
 \label{e:bi-orthogonal_normalization}
       \langle\alpha_m, v_n\rangle=\delta_{mn} \ .
\eea
The Keldysh expansion of the resolvent $R_L(\omega)=(L -\omega I)^{-1}$
then leads \cite{besson_quasi-normal_2025} to the asymptotic Keldysh QNM resonant expansion of the  scattered time-domain waveform  $u$ solving 
Eq. (\ref{e:wave_eq_1storder_u_tau})
       \bea
   \label{e:Keldysh_QNM_expansion_conclusions}
   u(\tau,x) &\sim& \sum_n a_n v_n(x)  \;e^{i\omega_{n}\tau} \ ,
   \eea
 where the coefficient $a_n$ (here  $\langle \alpha, v\rangle$ denotes the dual pairing
 between $\alpha\in {\cal B}^*$ and $v\in {\cal B}$) is
\bea
   a_n =
   \langle\alpha_n, u_0\rangle \ ,
   \eea
 with $u_0$ the initial data for the scattering evolution
(see \cite{besson_quasi-normal_2025} for the precise meaning of ``$\sim"$). Such Keldysh expansion is however invariant under the rescaling: $v_n\to f \; v_n$, $a_n\to 1/f \;a_n$. Therefore coefficients $a_n$'s are not intrinsically well defined (in the absence of a choice for the  scalar product). However, the product $ a_n v_n(x)$
is indeed invariant~\cite{Ansorg:2016ztf}, thus leading to 
   \bea
    \label{e:Keldysh_QNM_expansion_conclusions}
   u(\tau,x) \sim 
   \sum_n{\cal A}_n(x) \;e^{i\omega_{n}\tau}  \quad ,  \quad {\cal A}_n(x) \equiv  a_n v_n(x) = \langle\alpha_n, u_0\rangle v_n(x),
   \eea
recovering the Lax-Phillips asymptotic resonant expansion of a scattered field~\cite{LaxPhi89,dyatlov2019mathematical,hintz_quasinormal_2021}.

Our main interest is to monitor the activation of QNMs in time, aiming at relating
such QNM activation evolution to the qualitative features of the time-domain waveform. With
this aim we consider the expansion of $u$ in 
independent time-evolving modes ${\cal A}(\tau, x)$, as
\bea
\label{e:}
    u(\tau,x) \sim \sum_n {\cal A}_n(\tau, x)  \quad , \quad {\cal A}_n(\tau, x) \equiv {\cal A}_n(x) \; e^{i\omega_{n}\tau} \ .
   \eea
This decomposition motivates the following mode-quantities to be monitored in time:
\begin{itemize}
\item[i)] Modulus of the complex amplitude ${\cal A}_n(\tau, x)$:
\bea
\label{e:module_A}
|{\cal A}_n(\tau, x)| = |{\cal A}_n(x)\;e^{i\omega_{n}\tau}| =
|{\cal A}_n(x)|\;e^{-\mathrm{Im}(\omega_n)\tau} \ .
\eea

\item[ii)] {\em Relative modal weight} $W_n(\tau, x)$, defined as:
\bea
\label{e:W_n}
W_n(\tau, x) \equiv \frac{|{\cal A}_n(\tau, x)|}{\max_n|{\cal A}_n(\tau, x)|} = \frac{|{\cal A}_n(x)\;e^{i\omega_{n}\tau}|}{\max_n|{\cal A}_n(x)\;e^{i\omega_{n}\tau}|} \ .
\eea

\end{itemize}
The ultimate justification for these quantities is provided by the results in the next section. We focus here on these quantities evaluated at null infinity, at $x=x_{\scri^+}$, that 
we denote respectively as (the modulus of) ${\cal A}^{\infty}_n(\tau)\equiv{\cal A}_n(\tau, x_{\scri^+})$ and as
$W_n^{\infty}(\tau) =W_n(\tau, x_{\scri^+})$.

 \section{QNM expansion of Vishveshwara's waveform}
\begin{figure}[htp]
  \centering
  \includegraphics[clip,width=0.71\columnwidth]{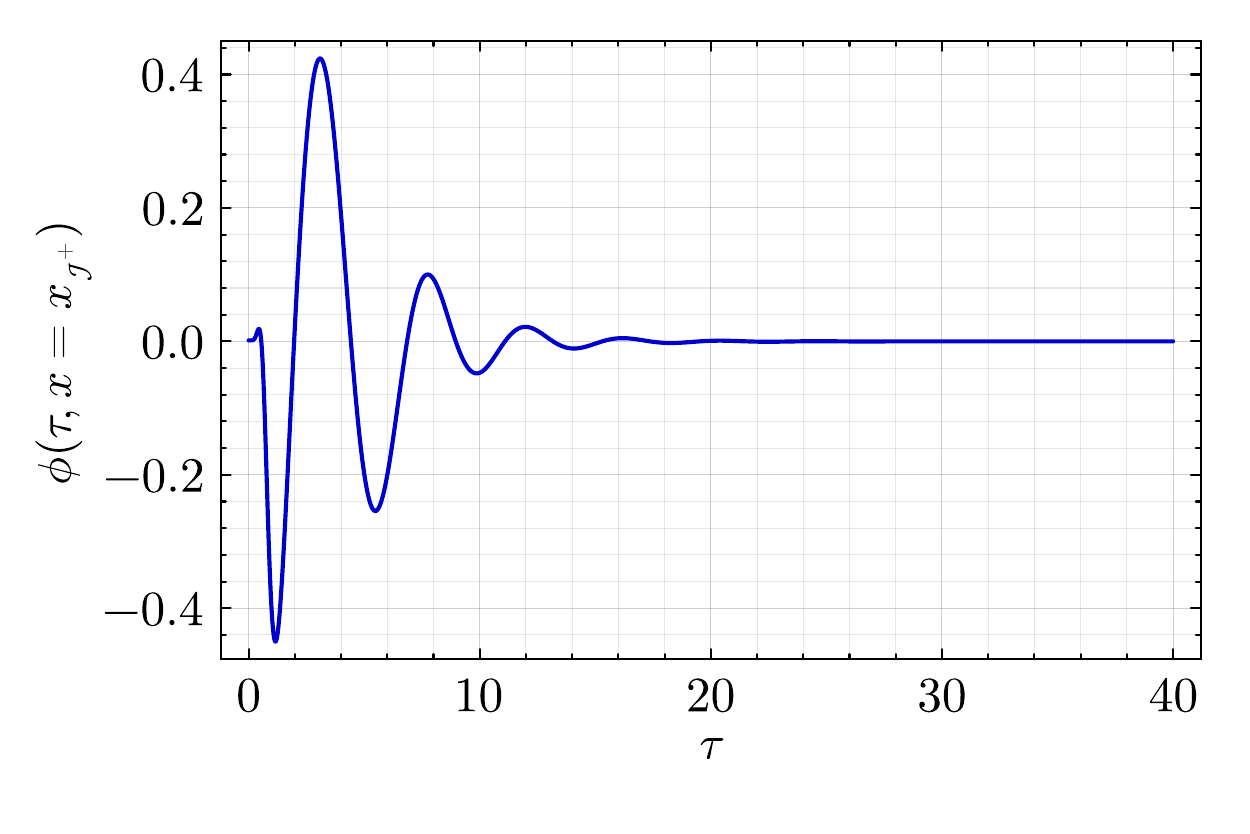}
  \caption{Vishveshwara-like waveform, result of the scattering of a Gaussian wavepacket by a Schwarzschild BH (Zerilli), calculated in a hyperboloidal slicing scheme (here $\lambda = 4M$).}
  \label{f:Vishveshwara_waveform}
\end{figure}

In the steps of \cite{vishveshwara1970scattering}, we study the scattering by a Schwarzschild BH of a 
($\ell=2$)  gravitational-perturbation wave-packet with Gaussian profile. For concreteness, we consider initial data 
$u_0= \left(\phi_0(x)= \exp\left(-\frac{1}{2}\frac{(x-x_0)^2}{\sigma^2}\right), \psi_0(x)=0\right)^t$. Rather than the Regge-Wheeler effective potential (odd perturbations) as in \cite{vishveshwara1970scattering}, we
consider first the Zerilli effective potential (even perturbations).
Fig. \ref{f:Vishveshwara_waveform} shows the resulting familiar Vishveshwara-like waveform (for an initial Gaussian profile with $\sigma=0.125$, $x_0=0.45$) evaluated at $\scri^+$. Note that we show the waveform in  hyperboloidal time $\tau$, in contrast to Fig. 3 in
\cite{vishveshwara1970scattering} showing a 
snapshot in $r^*$.

\begin{figure}[htp]
    \centering
    \subfloat[$\tau=0.6$]{
      \includegraphics[clip,width=0.4\columnwidth]{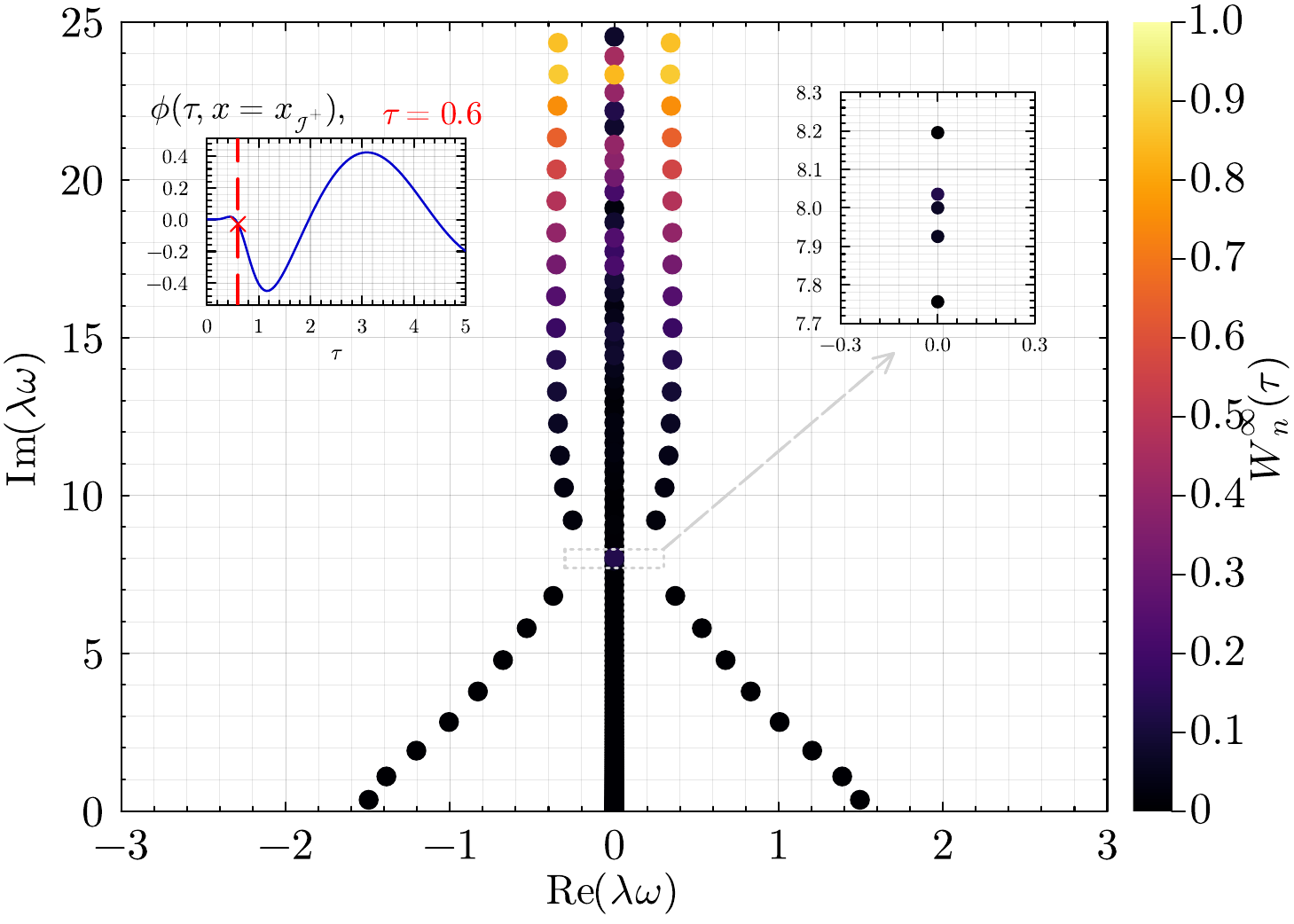}\label{snapshot_0.6}
    }
    \subfloat[$\tau=0.75$]{
      \includegraphics[clip,width=0.4\columnwidth]{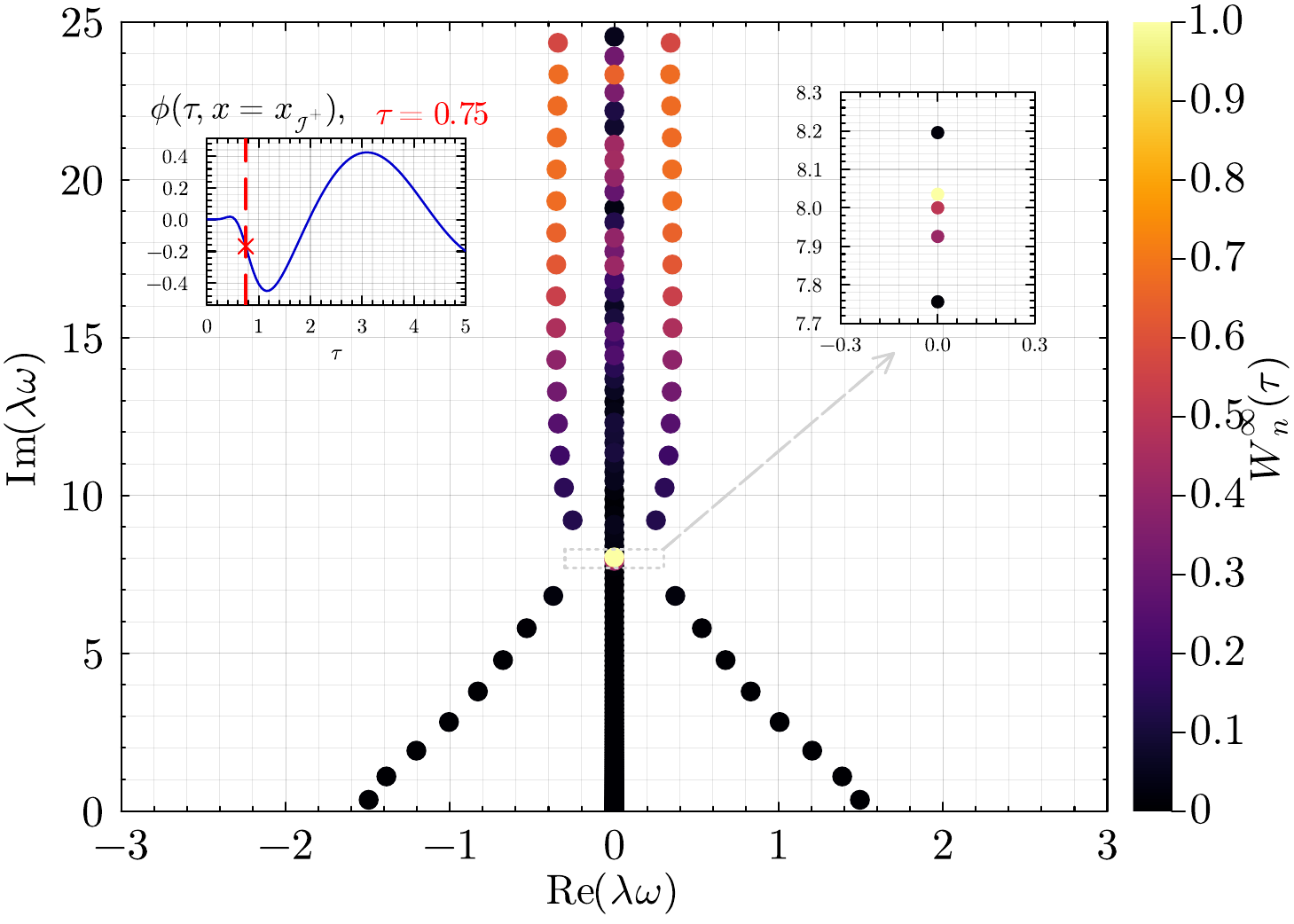}\label{snapshot_0.75}
    }
    \subfloat[$\tau=1.1$]{
      \includegraphics[clip,width=0.4\columnwidth]{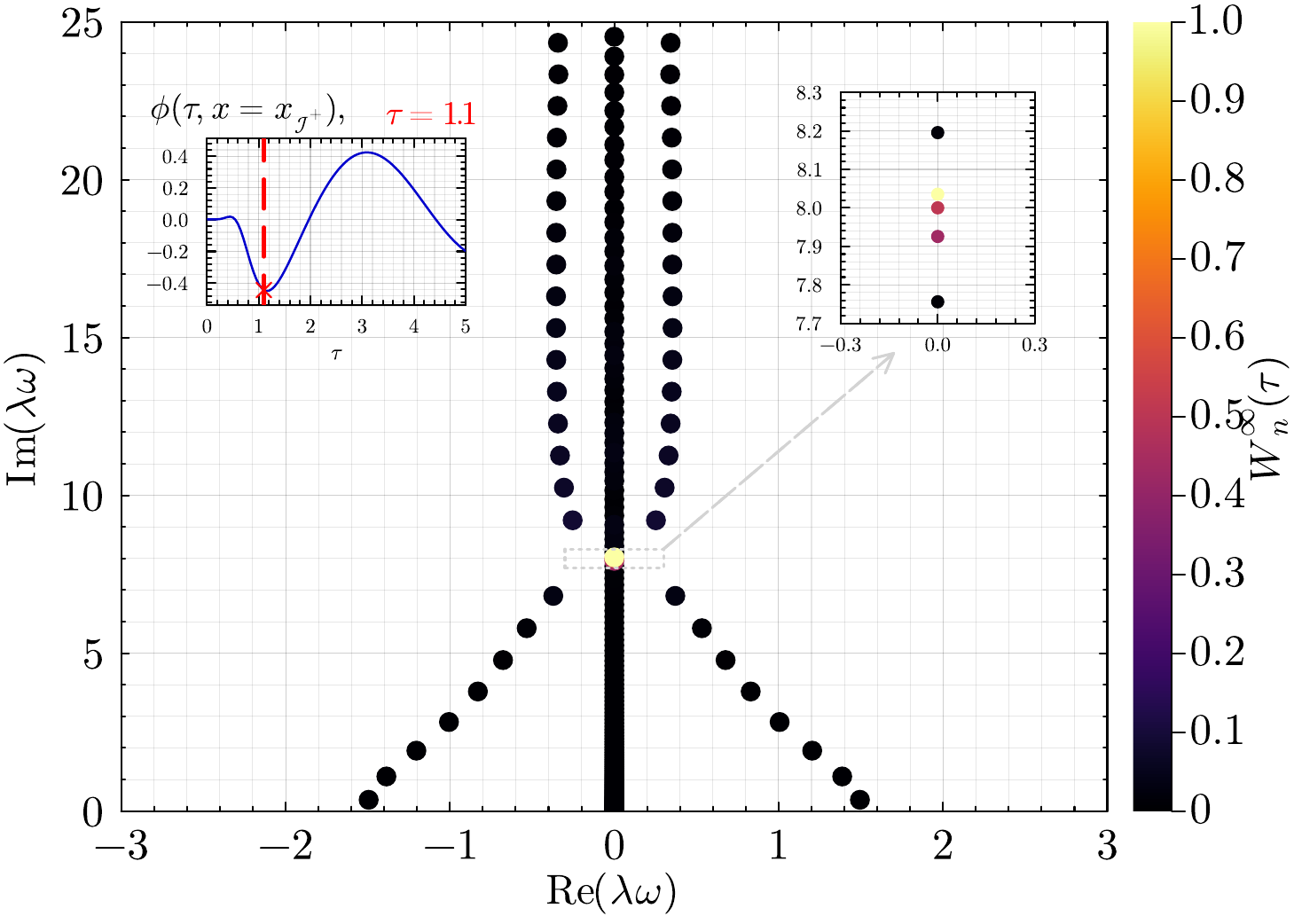}\label{snapshot_1.1}
    }
  
    \subfloat[$\tau=2.5$]{
      \includegraphics[clip,width=0.4\columnwidth]{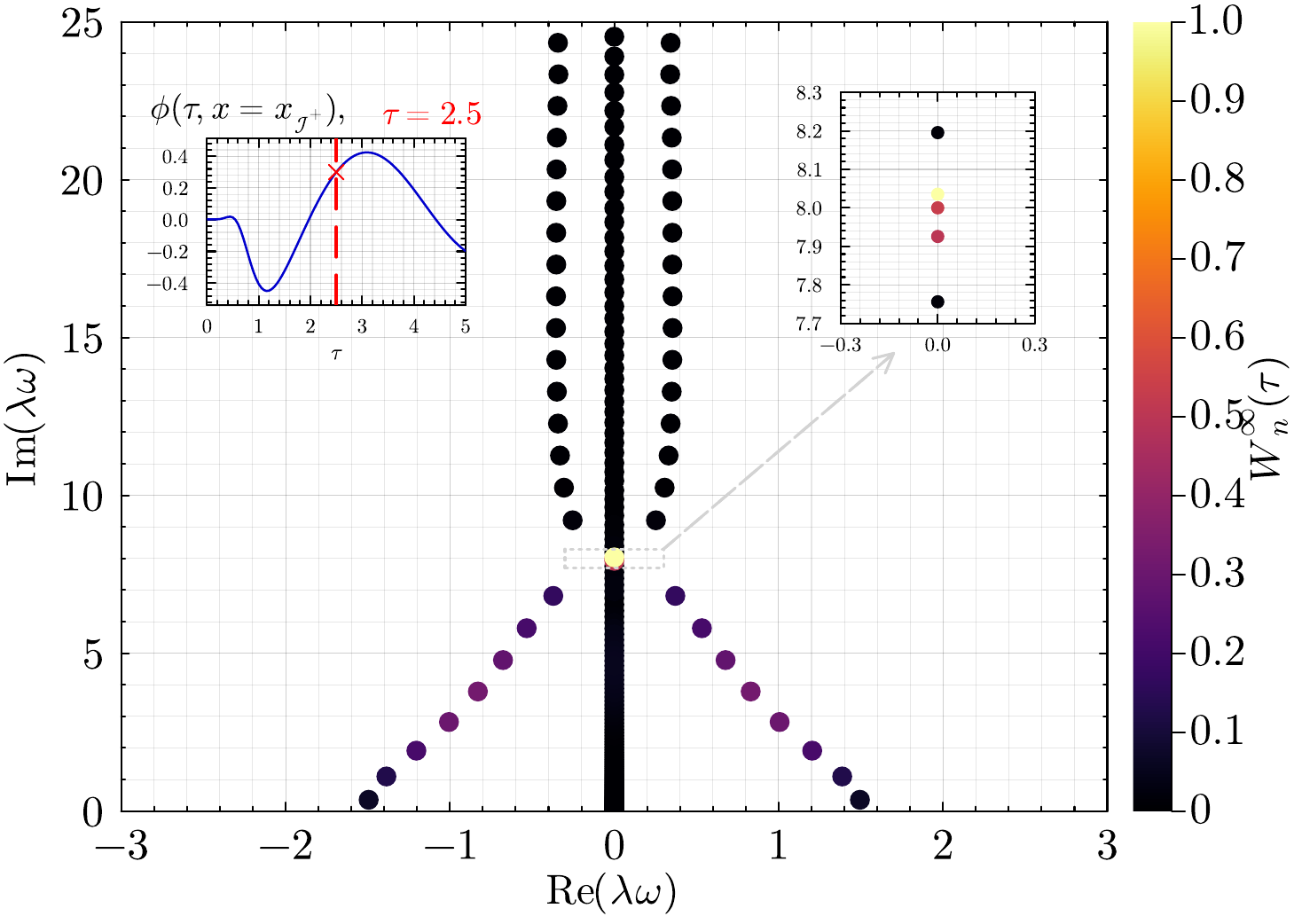}\label{snapshot_2.5}
    }
    \subfloat[$\tau=2.65$]{
      \includegraphics[clip,width=0.4\columnwidth]{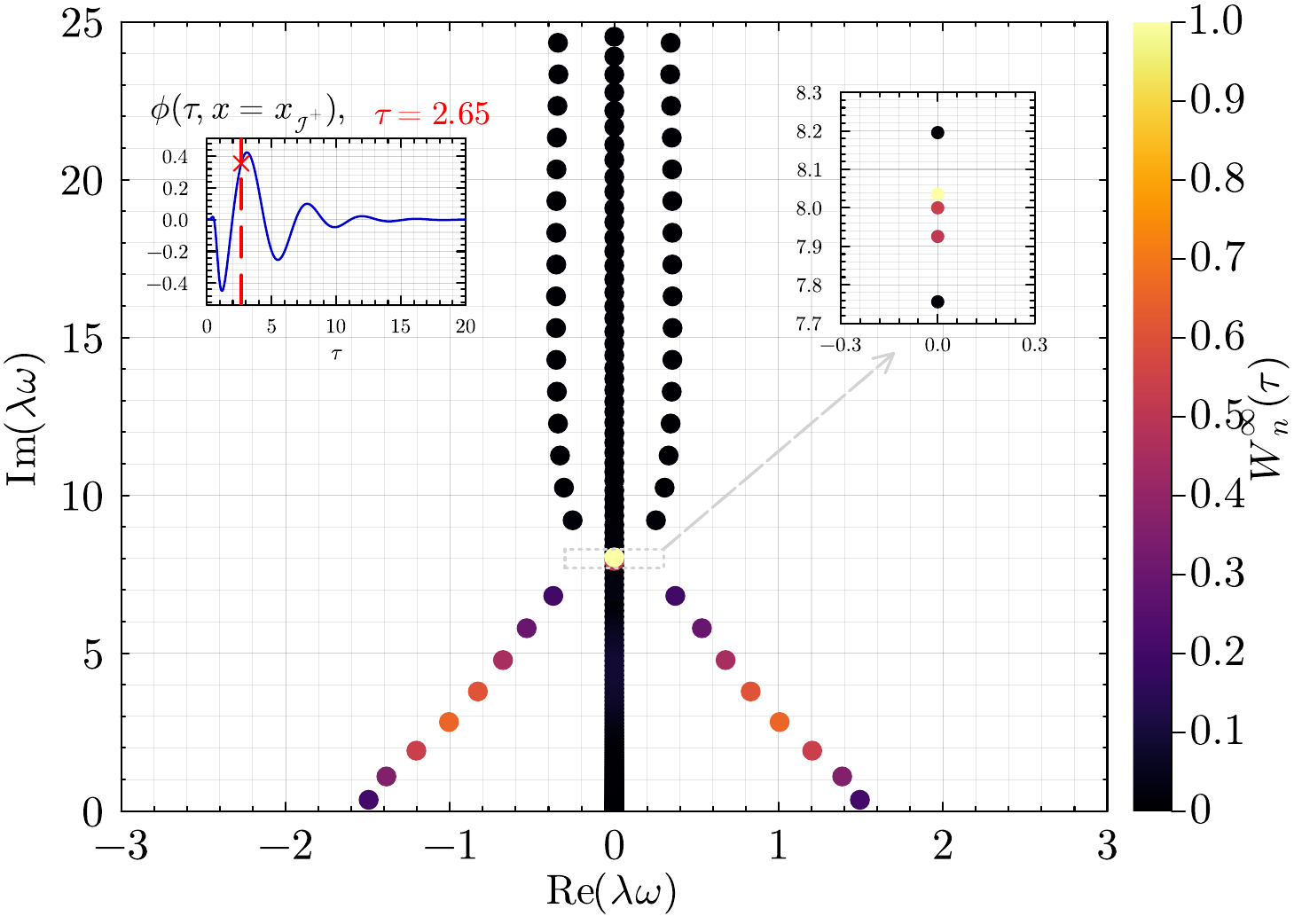}\label{snapshot_2.65}
    }
    \subfloat[$\tau=2.8$]{
      \includegraphics[clip,width=0.4\columnwidth]{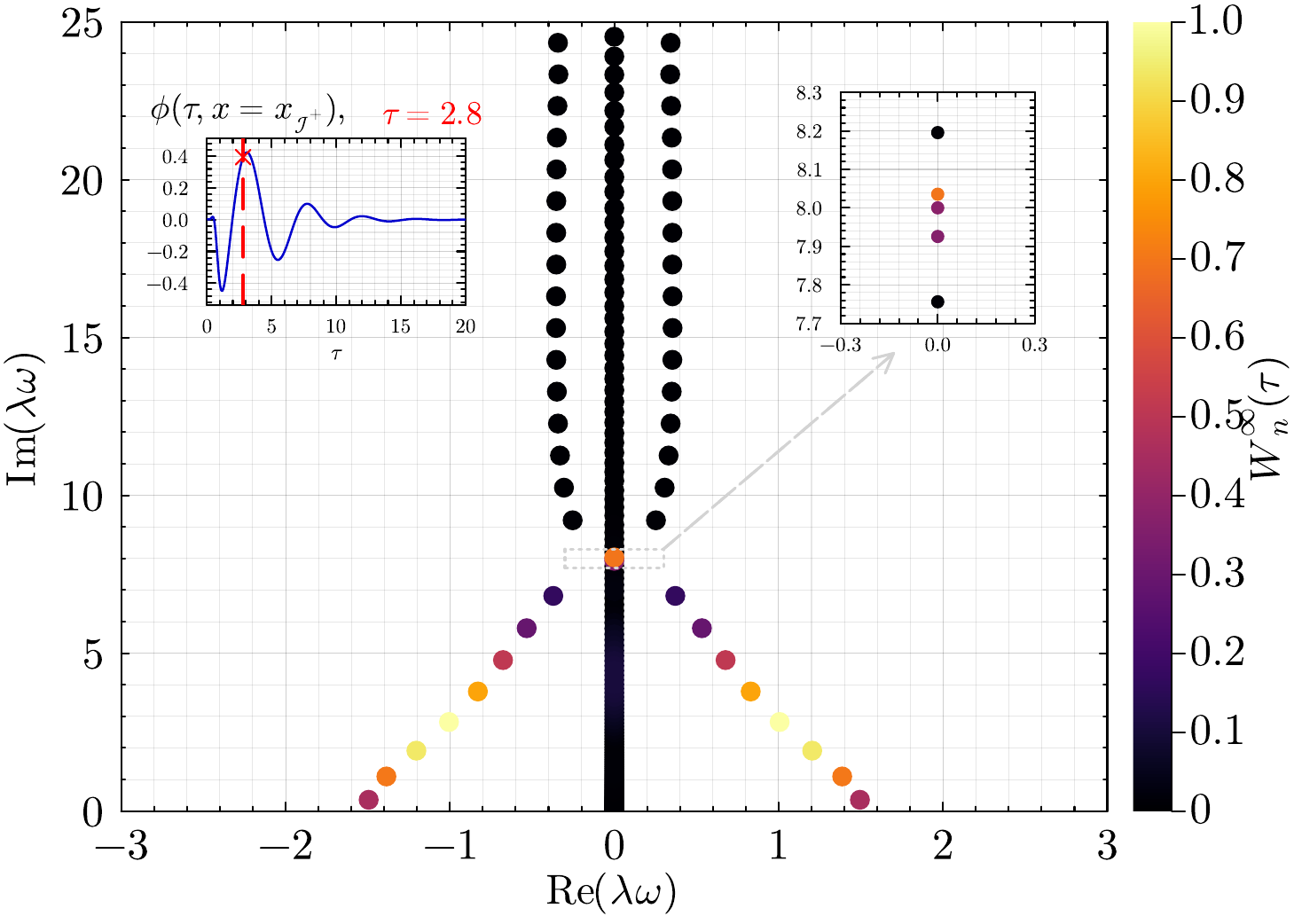}\label{snapshot_2.8}
    }
    
    \subfloat[$\tau=3.1$]{
      \includegraphics[clip,width=0.4\columnwidth]{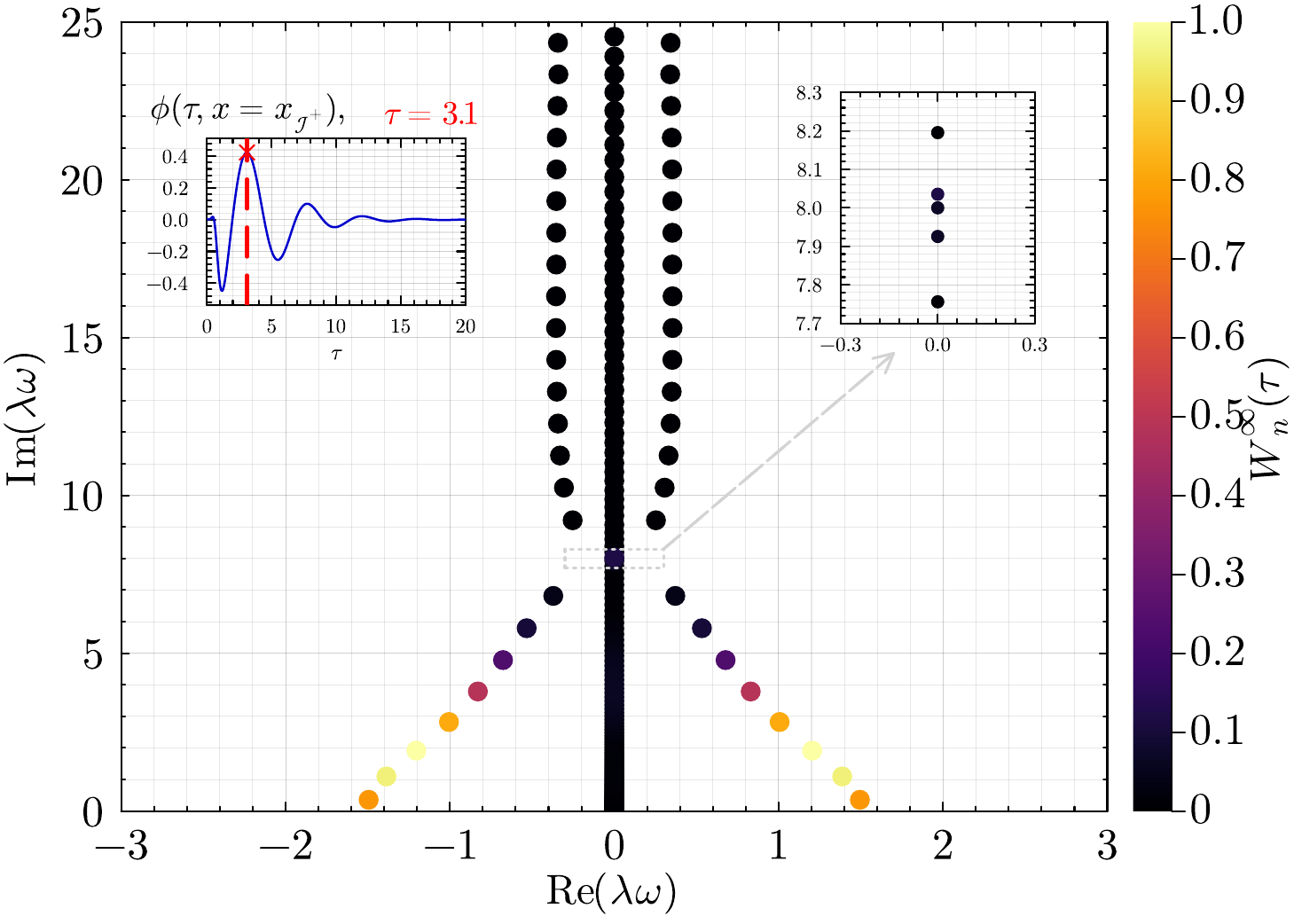}\label{snapshot_3.1}
    }
    \subfloat[$\tau=3.4$]{
      \includegraphics[clip,width=0.4\columnwidth]{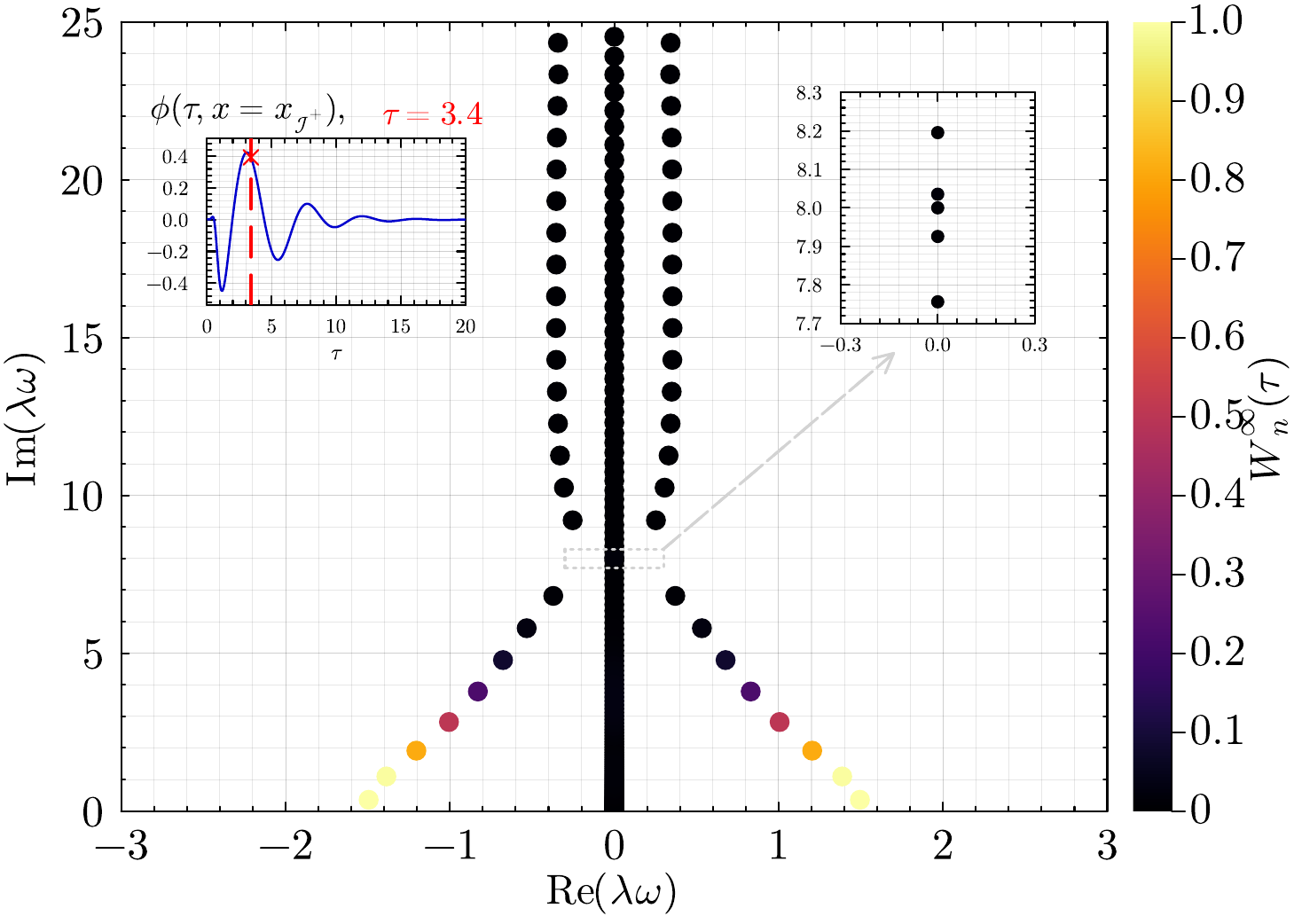}\label{snapshot_3.4}
    }
    \subfloat[$\tau=4.0$]{
      \includegraphics[clip,width=0.4\columnwidth]{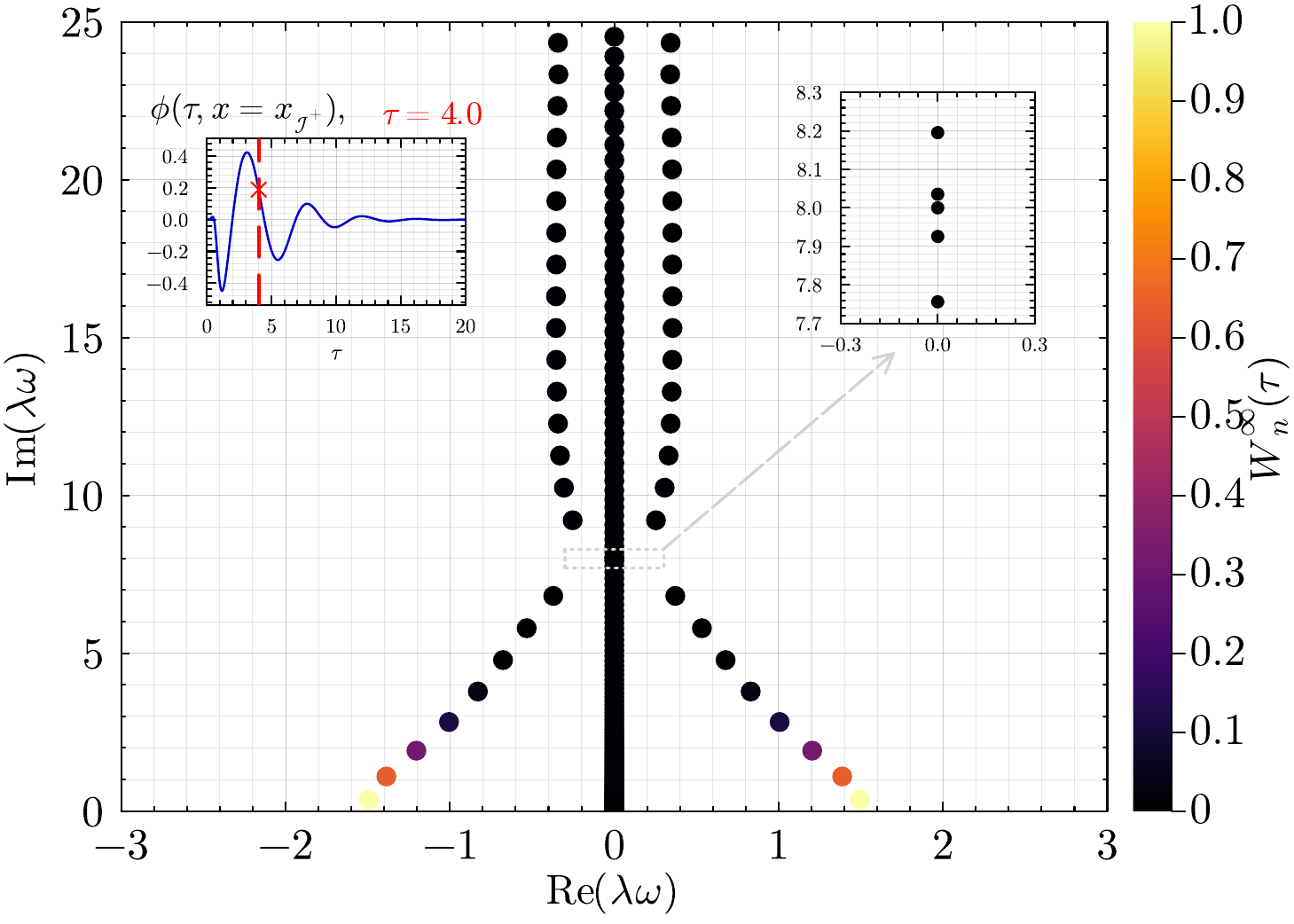}\label{snapshot_4.0}
    }
    
    \subfloat[$\tau=4.8$]{
      \includegraphics[clip,width=0.4\columnwidth]{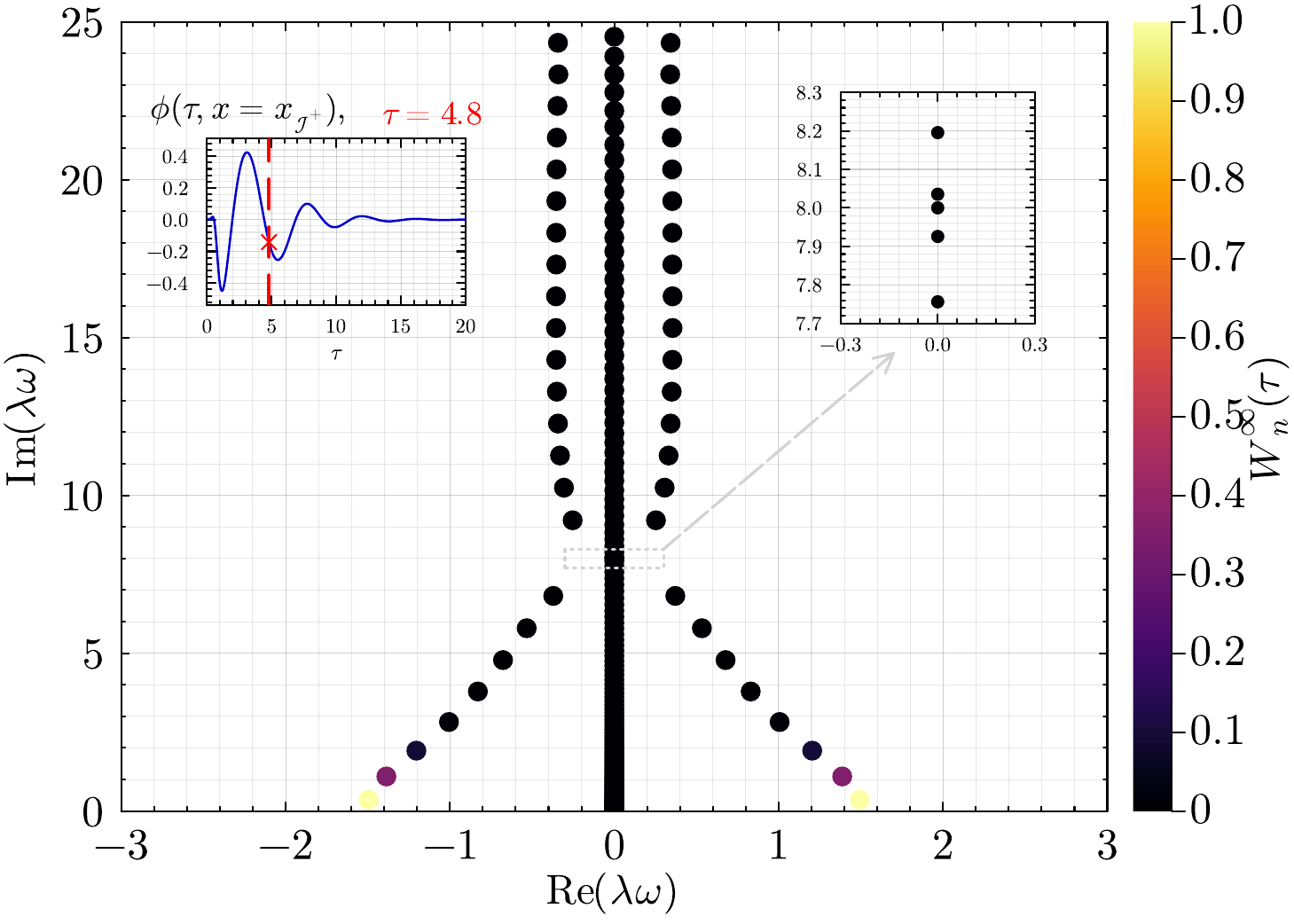}\label{snapshot_4.8}
    }
    \subfloat[$\tau=5.6$]{
      \includegraphics[clip,width=0.4\columnwidth]{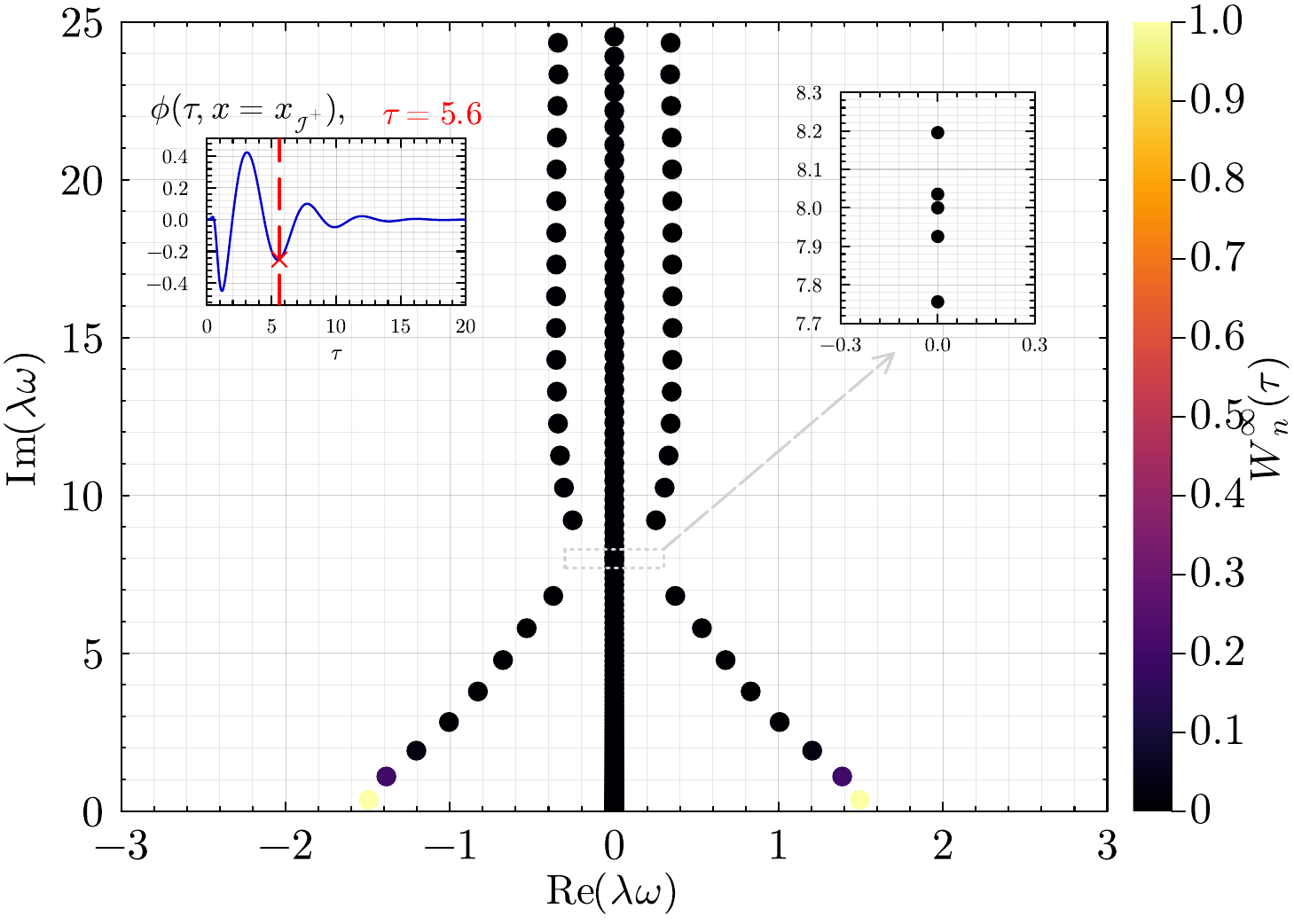}\label{snapshot_5.6}
    }
    \subfloat[$\tau=39.0$]{
\includegraphics[clip,width=0.4\columnwidth]{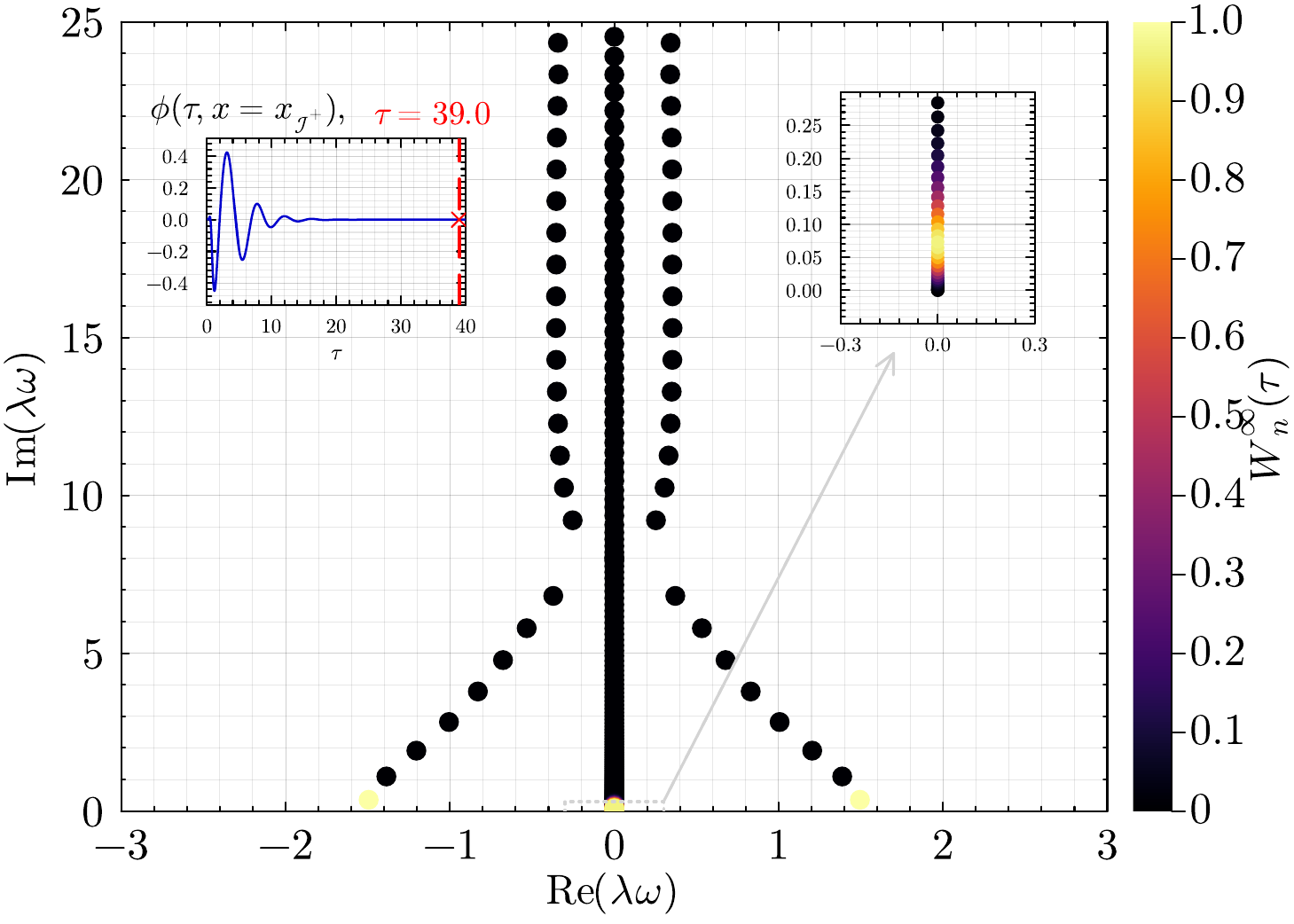}\label{snapshot_39.0}
    }
    \caption{Time activation of QNM modes and branch cut in the Schwarzschild BH (Zerilli case, $\lambda=4M$). Each snapshot displays the complex QNM frequencies and (the numerical approximation of) the branch cut, their activation 
  being monitored by the value of $W^\infty_n(\tau)$, that can be read from the colour scale. The corresponding time in the waveform is shown in an inset. Activated frequencies descend in the complex plane with time, with a `stagnation' phase  
  $\tau\approx 0.75-2.65$  in modes around the AS frequency (zoomed in an inset) and coinciding with the first peak. Then low overtones and the fundamental QNM dominate from the second peak and, from time $\tau \approx 5.6$, the mode activation remains stable and dominated by the fundamental QNM, until very late times 
  ($\tau\approx 39.0$) dominated by the branch cut (tails) close to the origin.
  }
  \label{f:snapshots}   
  \end{figure}

\begin{figure}[htp]
  \centering
  \includegraphics[clip,width=0.6\columnwidth]{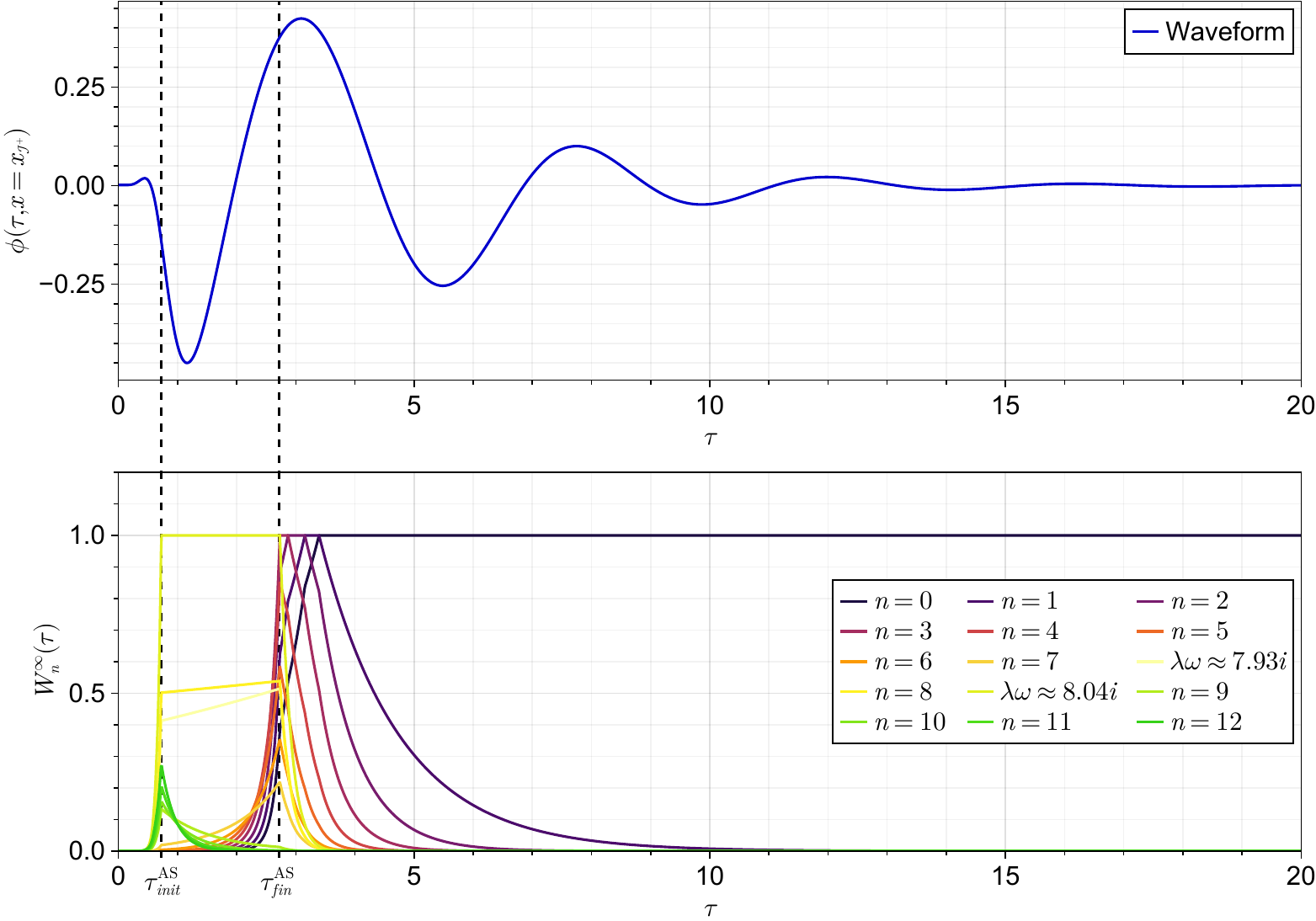}\label{u0:u0}
  \caption{Evolution of $W^\infty_n(\tau)$. Three phases can be
  identified: i) a ``prompt phase" (greens), ii) a ``prompt-to-ringdown phase" (yellows)  
  and iii) a ``ringdown phase" (red-violet-black), separated
  by dashed lines at 
  $\tau^\mathrm{AS}_{\mathrm{init}}=0.72$ and at $\tau^\mathrm{AS}_{\mathrm{fin}}=2.72$. Late tails have been excluded.}
  \label{f:W}
\end{figure}

We start by studying the activation of QNMs in time, by monitoring  $W^\infty_n(\tau)$. The results 
are presented in Fig. \ref{f:snapshots}.  We display
the QNM frequencies at different times $\tau$, with colours reflecting their activation and corresponding to the value $W^\infty_n(\tau)$. Each snapshot 
shows an inset with the time evolution along the waveform, to ``correlate" the QNM activation
with the waveform qualitative features.
We describe the sequence of time-snapshots in Fig. \ref{f:snapshots}: 
\begin{itemize}
    \item[1.] {\em Prompt signal}. Before $\tau\approx 0.8$ ($\lambda=4M$), it is dominated by high overtones (the upper part of the branch cut is also activated). As time goes by frequencies
    smoothly go down.

    \item[2.] {\em First (dominant) peak and AS frequency}. Around $\tau\approx 0.8$ the descent of frequencies is drastically altered by the sudden activation of
    modes
    close to the AS frequency $\lambda\omega_{AS}=8i$. This AS phase
    fully dominates the first peak, in $\tau\approx 0.75- 2.65$, the AS frequencies only extinguishing completely at the apex of the second peak, at $\tau\approx 3.1$. 

\item[3.] {\em Second (dominant) peak and ``fundamental mode and lowest-overtone onset"}. QNM overtones below the AS frequency start being activated shortly before the apex of the second peak, around $\tau\approx 2.5$, and smoothly descend in the complex plane. The region around the apex of the second peak ($\tau\approx 2.8 - 3.4$) is dominated by the first overtones and the fundamental mode. Shortly after the peak ($\tau\approx 3.4$)
the signal is dominated by the fundamental QNM and the first overtone, the latter extinguishing by  $\tau\approx 4.8$.

\item[4.] {\em Fundamental mode and late time tails.} After
 $\tau\approx 4.8$ the signal is fully dominated by the fundamental mode with the snapshot at $\tau\approx 5.6$ providing 
 the picture until very late times, when power-law tails enter the 
 scene,
 dominating the signal around  $\tau\approx 39.0$.

\end{itemize}
Some remarks are in order:
\begin{itemize}
\item[i)] Although the notion of `ringdown onset' does not 
make proper sense in a linear scattering such as this one, Fig. \ref{f:snapshots} suggests that the expected BH ringdown decay phase, controlled by the fundamental QNM and
first overtones, starts not earlier than the second peak. The spectral-domain analysis suggests a specific
time (or at least an earliest bound) to the `ringdown onset',
that in the waveform in Fig. \ref{f:Vishveshwara_waveform} 
is given by the
second dominant peak.

\item[ii)] The `AS stagnation' during the first dominant peak is a phenomenon involving the AS QNM and the neighbouring branch cut\footnote{We note that, in the matrix-approximants of the operator $L$ at a given Chebyshev grid resolution, the branch cut is approximated as a finite number of (non-convergent) eigenvalues along the imaginary axis. See details in 
\cite{Ansorg:2016ztf,PanossoMacedo:2018hab,Jaramillo:2020tuu}.}. 
An inset at each snapshot  displays a zoom around the AS frequency exhibiting that, though the AS QNM is indeed strongly excited,
the maximally activated frequency is actually a nearby one in the branch cut. When changing the grid resolution/initial data, the realised branch cut frequencies change, but the phenomenon remains: both the AS QNM and close branch cut points get activated.

\item[iii)] 
The late-time branch cut activation is shown in panel (\ref{snapshot_39.0}) (see the zoom in the inset) when, as the fundamental QNM decays, the power-law tail decay becomes dominant in the signal and drives the late time behaviour in quantitative accordance with Price law.

\item[iv)] The qualitative features in Fig. \ref{f:snapshots} remain unchanged when the Gaussian wavepacket parameters are varied or, more generally, 
if modifying
the pattern of the initial data wavepacket, as long as its `width' is no larger than the lengthscale of the BH.
\end{itemize}
A more quantitative illustration of the mode activation is presented in Fig. \ref{f:W}, where $W^\infty_n(\tau)$ is plotted in time for several QNMs and
branch cut modes, the latter around the AS frequency. From the very 
definition of $W^\infty_n(\tau)$, at any time $\tau$ we have
$W^\infty_n(\tau)=1$ for some mode (this seems to fail at very early times simply because high overtones are not
displayed). The relevant information is in the colour change 
along the line $W^\infty_n(\tau)=1$, that identifies the dominating mode at time $\tau$, and the quantitative relative contribution of other active modes.  
\\
From Fig. \ref{f:W} we identify clearly three dynamical stages, separated by dashed lines: a ``prompt phase" dominated by overtones above the AS frequency (greens), a ``prompt-to-ringdown phase" controlled by (QNM and branch cut) modes around the AS frequency (yellows), and a ``ringdown phase" with low overtones and the fundamental QNM (red-violet-black). 

\subsection{`AS stagnation': enhancement of modes near AS frequencies in the even-parity case}
\label{s:AS_stagnation}
\begin{figure}[htp]
    \centering
    \subfloat[Zerilli]{
        \includegraphics[clip,width=0.55\columnwidth]{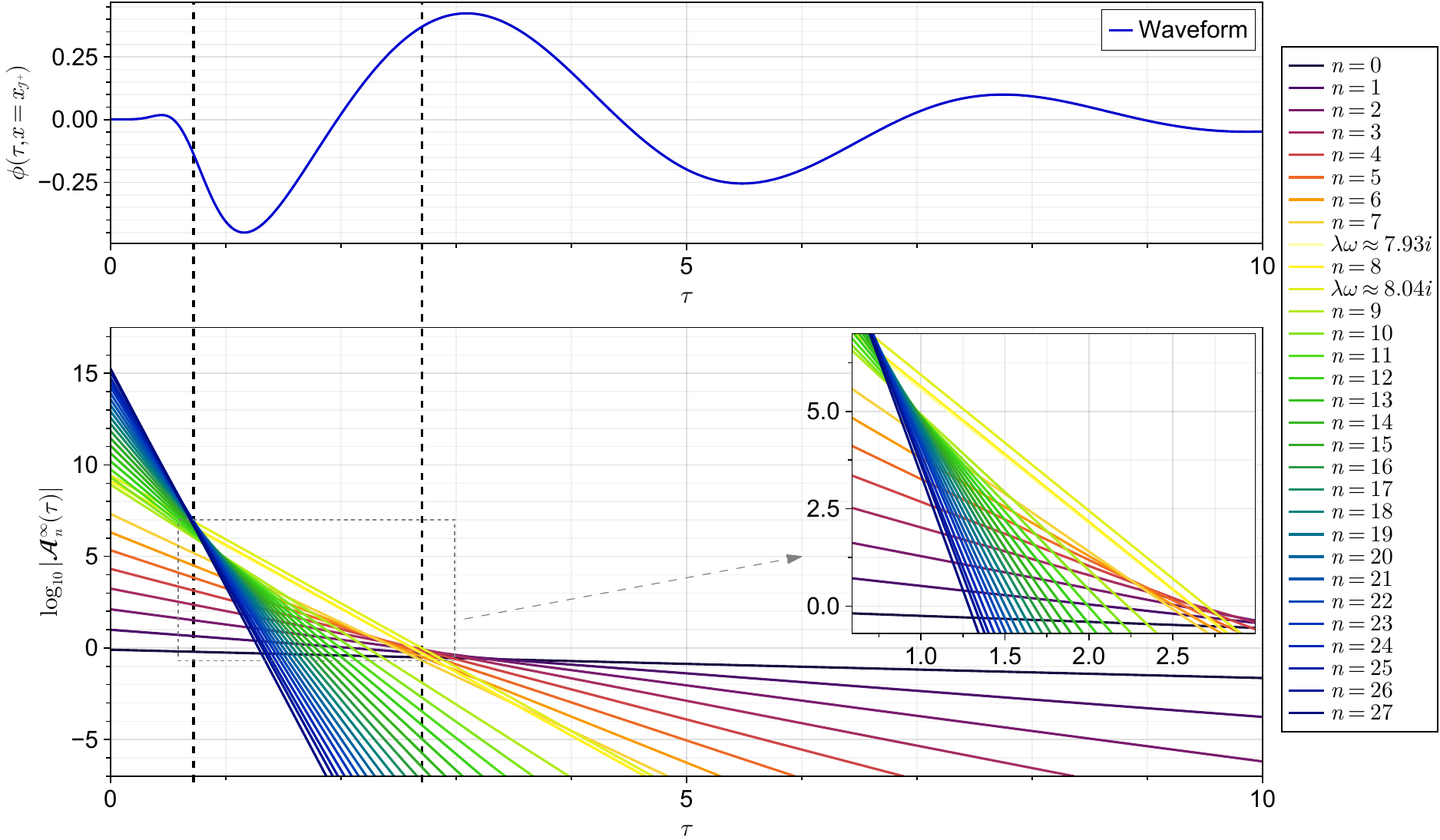}\label{panel:Z:enveloping}
    }
    \subfloat[Regge-Wheeler]{
      \includegraphics[clip,width=0.55\columnwidth]{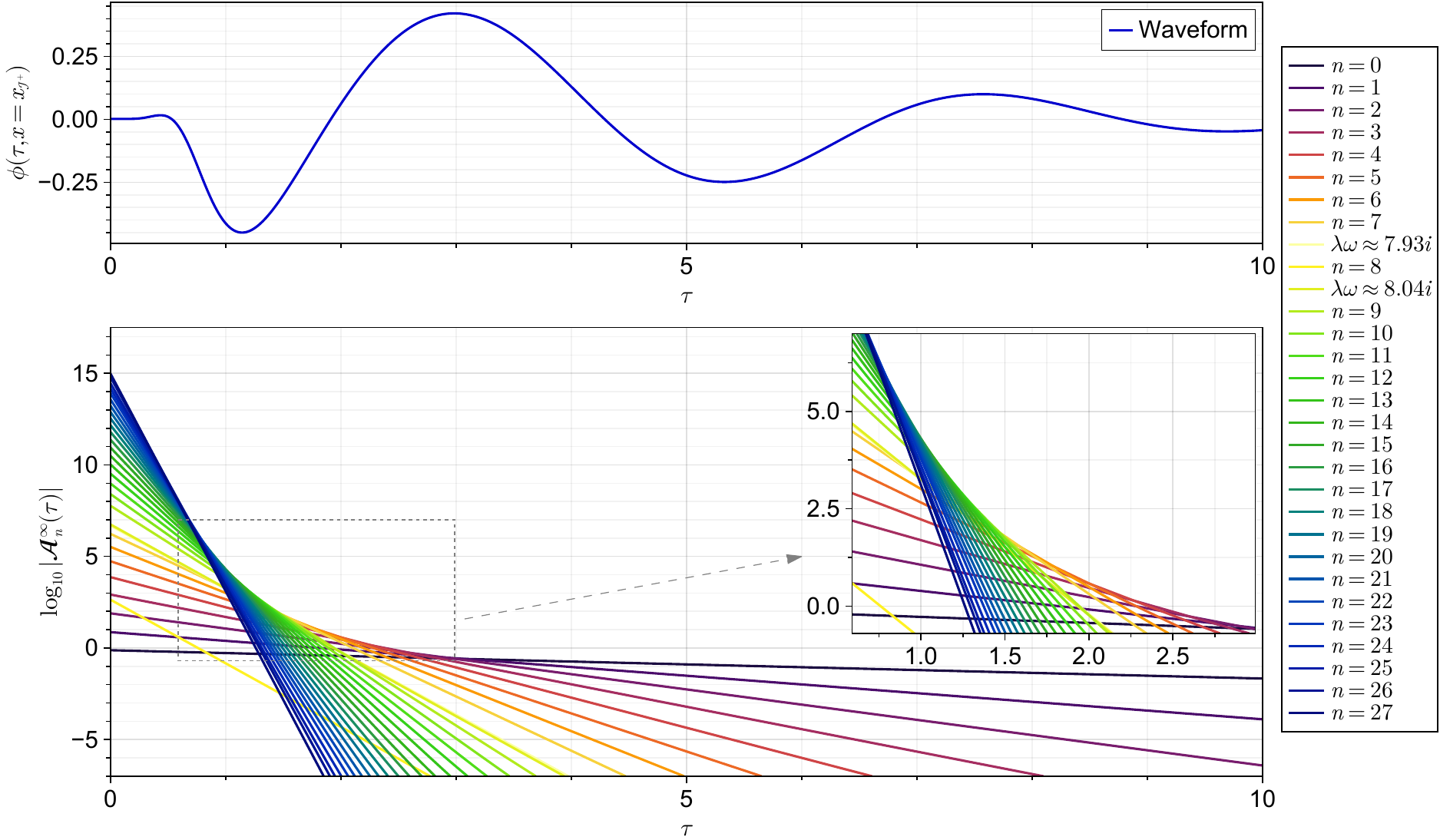}\label{panel:RW:enveloping}
    } 
    \caption{Evolution of $|{\cal A}^\infty_n(\tau)|$ in Schwarzschild ($\lambda=4M$). In a logarithm representation, the set of straight QNM lines 
    produce a smooth ``envelope" reflecting the `smooth'  activation (as $\omega_n$ changes in $n$) of QNMs. The presence of a ``secant-to-the-envelope" is the ``smoking gun" of the AS stagnation phenomenon, produced by an anomalously enhanced activation of modes close
     to the AS frequency. The left panel 
     shows its presence in the Zerilli case (even perturbations), while the right panel  
     shows its absence for Regge-Wheeler (odd perturbations).}
     \label{f:enveloping}
\end{figure}

We focus now on the ``prompt-to-ringdown phase", controlled by modes around the AS frequency, that is `correlated' to the first peak up to a time close to the second peak apex. Specifically,
we stress two structural features that seem to play a  
relevant role:

\begin{itemize}
\item[a)] {\em Branch cut}: In point ii) above we have remarked that modes in the branch cut are indeed activated and are even dominant over the AS QNM during the `AS stagnation'. The existence of such a branch cut critically depends on the
considered asymptotic flatness, suggesting a role of spacetime
asymptotics. Interestingly, if monitoring
 the BH horizon ${\cal H}$ instead of null infinity, i.e.  $W^{\cal H}_n(\tau) \equiv W_n(\tau,x_{\Hor})$ instead of 
$W^{\infty}_n(\tau)$, the  `AS stagnation' is absent. 
We retain the indication that the presence of a branch cut (and therefore of asymptotically flat asymptotics) plays a key role in the phenomenon.

\item[b)] {\em QNM eigenvalue}: The `AS stagnation' observed in the even-parity case is absent
for odd-parity perturbations, that is, for the Regge-Wheeler effective potential (actually the case studied in \cite{vishveshwara1970scattering}). From a spectral perspective, we recall the remarkable QNM isospectrality between the even-parity (Zerilli) and the odd-parity (Regge-Wheeler) cases~\cite{ChaDet75,Chandrasekhar:579245}.
But there is a key mismatch:  whereas the Zerilli even-parity case does possess an AS QNM (together 
with a `total transmission mode' (TTM) from the BH horizon; cf. e.g. \cite{van2000analytic,Berti:2009kk}), 
the Regge-Wheeler problem does not have a proper AS QNM eigenvalue.
We retain here the presence of an actual AS QNM in `AS stagnation'.
\end{itemize}
In order to better illustrate the referred difference between the even-parity and odd-parity cases, with regard to the `AS stagnation' phenomenon, we directly monitor  
$|{\cal A}^\infty_n(\tau)|$ in Eq. (\ref{e:module_A}), rather
than $W_n^\infty(\tau)$. In
Fig. \ref{f:enveloping} we show, for both Zerilli and 
Regge-Wheeler cases, the logarithm of $|{\cal A}^\infty_n(\tau)|$ against time 
$\tau$ for QNMs and branch cut modes around the AS frequency.
The difference between both cases is apparent: whereas in the
Zerilli case the AS (QNM and branch cut) modes (in yellow)
produce a ``secant" in the smooth ``envelope" of all other QNM straight lines, such a secant is absent for Regge-Wheeler.
The presence of such a ``secant-to-the-envelope" is a `graphical'
smoking gun of the occurrence of the `AS stagnation' phenomenon, their intersection
points determining the ``prompt-to-ringdown" phase. 

The presence of the ``secant-to-the-envelope" in the left panel of Fig. \ref{f:enveloping} is a consequence of the relative higher
value of $|{\cal A}^\infty_n(0)|=|{\cal A}_n(x_{\scri^+})|$
for AS frequencies, as compared with the neighbouring modes, that raises
the AS-frequencies ``straight lines" over the ``envelope".
That is, the activation of AS frequencies get `anomalously' 
enhanced at null infinity, with respect to nearby modes. 
As commented in point iv) above, this activation enhancement feature 
does not depend on the chosen initial data, a key point for the discussion below.

\subsection{Excitation ratio between even and odd-parity perturbations }
\begin{figure}[htp]
    \centering
    \subfloat[Quasinormal modes]{
      \includegraphics[clip,width=0.5\columnwidth]{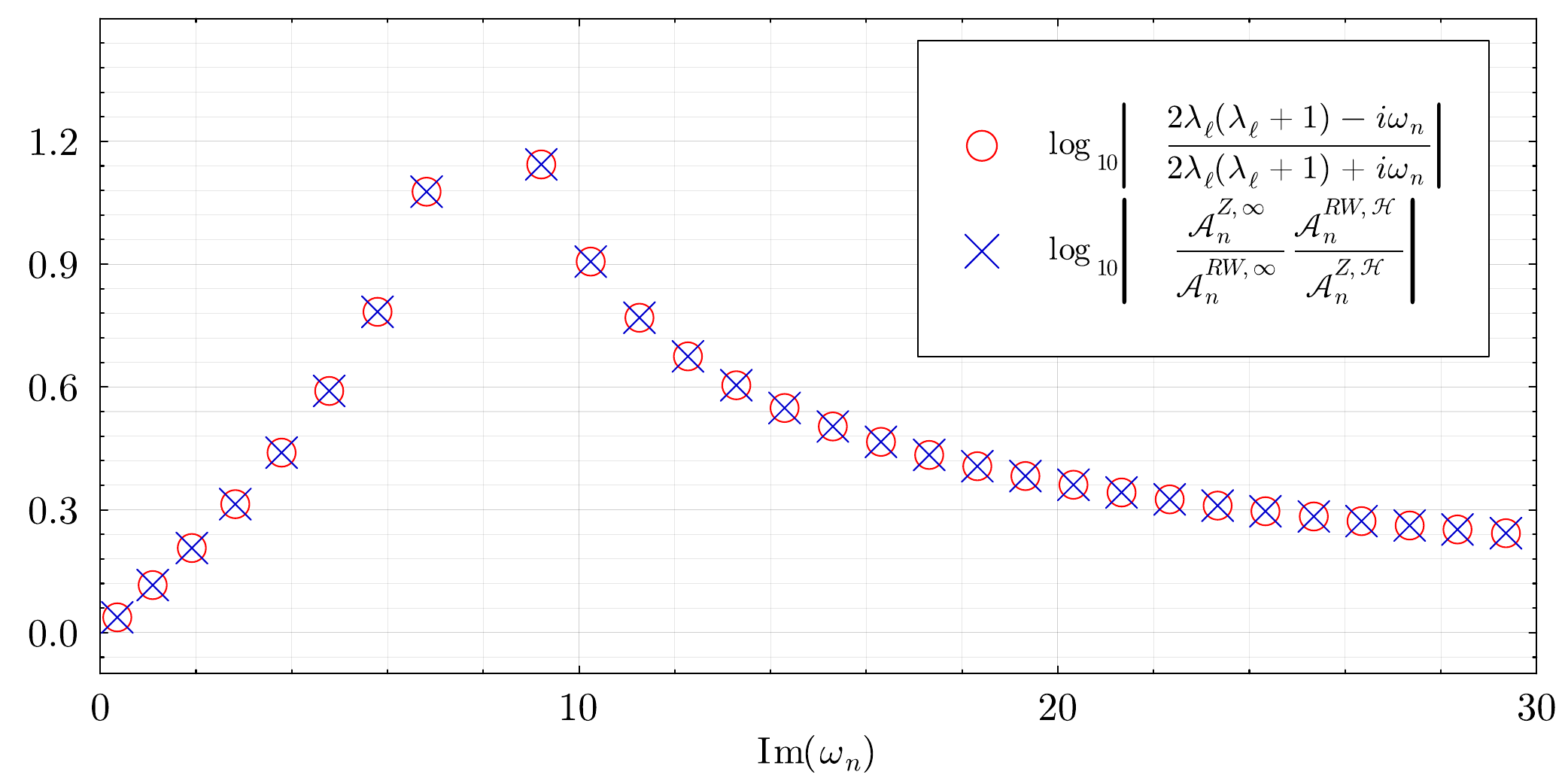}\label{CD:QNMs}
    }
    \subfloat[Branch-cut eigenvalues]{
      \includegraphics[clip,width=0.5\columnwidth]{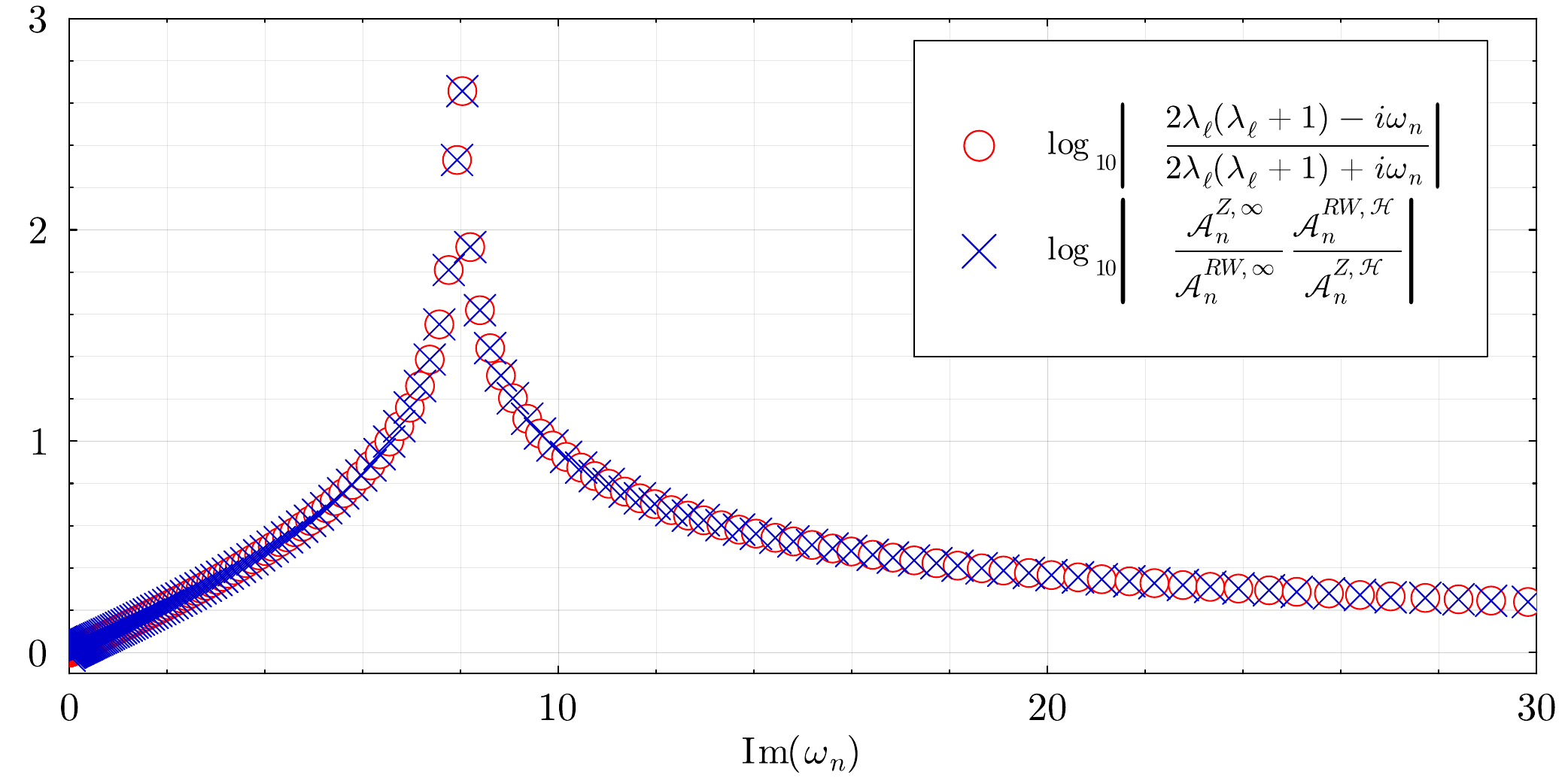}\label{CD:branch-cut}
    }
    \caption{Assessment of the ratio $\left|B^{\mathrm{Kel,even}}_n/B^{\mathrm{Kel,odd}}_n\right|$, for $\ell=2$. For each $n$, we consider
    the corresponding frequency $\omega_n$, both for QNMs (left panel)
    and branch-cut modes (right panel). The $x$-axis 
    corresponds to $\mathrm{Im}(\omega_n)$ (taken as a convenient parametrization for the plot), whereas in the $y$-axis 
    we represent the $\log_{10}$ of both $\left|B^{\mathrm{Kel,even}}_n/B^{\mathrm{Kel,odd}}_n\right|$
     (red circles) and of the modulus of the expression in Eq. (\ref{e:excitation_factors}) (blue crosses). The agreement 
    for both QNMs and branch cut is extremely good (the red circle at the AS frequency
    is not shown since Eq.(\ref{e:excitation_factors}) diverges), confirming
    the accuracy and robustness of the hyperboloidal implementation. }
    \label{f:CD_exact}
\end{figure}

The comparison between the
left and right panels in Fig. \ref{f:enveloping} also indicates
the enhancement of the AS frequencies in the even-parity 
case as compared to odd-parity perturbations.  Indeed,  it is known that the relative excitation of
even and odd-parity modes in Schwarzschild follows a simple law,
intimately related to the QNM-isospectrality referred 
above (actually a direct consequence of the underlying  Darboux transformation relating both sectors).
In the language of QNM excitation factors $B_n$,
that capture the QNM excitability  independently of the source
 (cf. e.g. \cite{ZhangBertiCardoso13} and references therein), the following `Darboux' relation holds \cite{ChaDet75, Leaver85, Chandrasekhar:579245}.
\bea
\label{e:excitation_factors}
\frac{B^{\mathrm{even}}_n}{B^{\mathrm{odd}}_n}=\frac{B^{\mathrm{even}}(\omega_n)}{B^{\mathrm{odd}}(\omega_n)}=
\frac{2\lambda_\ell(\lambda_\ell + 1) - 3i\omega_n}{2\lambda_\ell(\lambda_\ell + 1) + 3i\omega_n} \ ,
\eea
where $\lambda_\ell = (\ell - 1)(\ell + 2)/2$, for a gravitational angular $\ell$-mode. Although the notion of excitation factor
$B_n$ does not emerge naturally in the (Keldysh) framework,
from the resonant expansion in Eq. (\ref{e:Keldysh_QNM_expansion_conclusions}) we can introduce 
a number $B^{\mathrm{Kel}}_n$ capturing the excitation of the
mode $n$ (now either QNM or branch-cut mode), but independent
of the initial data, by considering
\bea
\label{e:Kel_n}
B^{\mathrm{Kel}}_n = \frac{{\cal A}_n(x_{\scri^+})}{{\cal A}_n(x_{\cal H})} \ ,
\eea
where $x_{\cal H}$ denotes the coordinate of the BH horizon and the ratio eliminates the dependence on the initial data $u_0$.
In Fig. \ref{f:CD_exact} we plot the ratio 
$|B^{\mathrm{Kel},\mathrm{even}}_n/B^{\mathrm{Kel},\mathrm{odd}}_n|$ for both QNMs and branch points, comparing it
with the corresponding evaluation of the right-hand-side in Eq. (\ref{e:excitation_factors}). The agreement is very accurate both for QNMs and branch-cut modes, the latter being particularly remarkable since branch-cut eigenvalues are approximants
of a continuum structure.  

\begin{figure}[htp]
    \centering
    \subfloat[Quasinormal modes]{
      \includegraphics[clip,width=0.5\columnwidth]{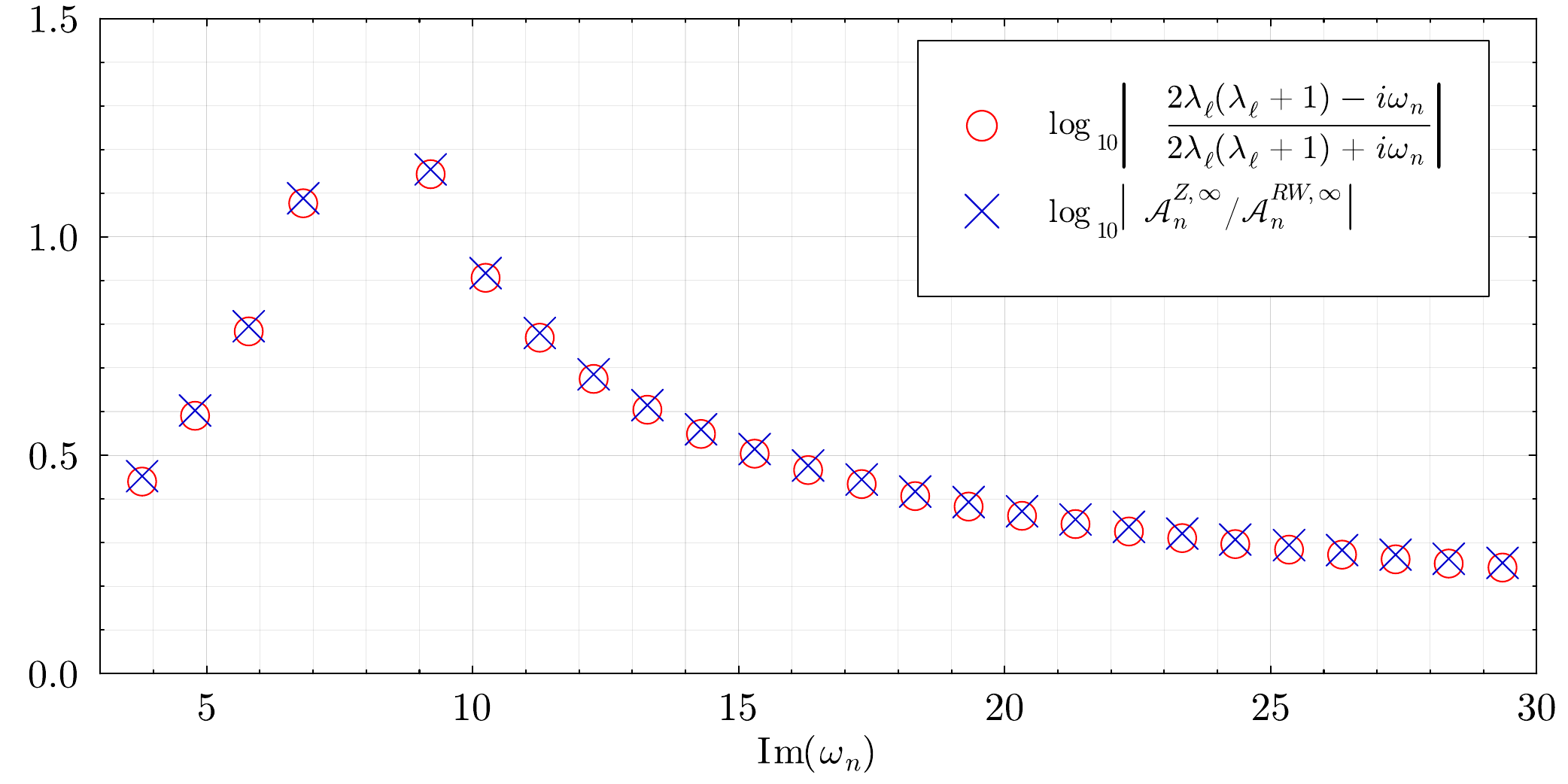}\label{CD:QNMs}
    }
    \subfloat[Branch-cut eigenvalues]{
      \includegraphics[clip,width=0.5\columnwidth]{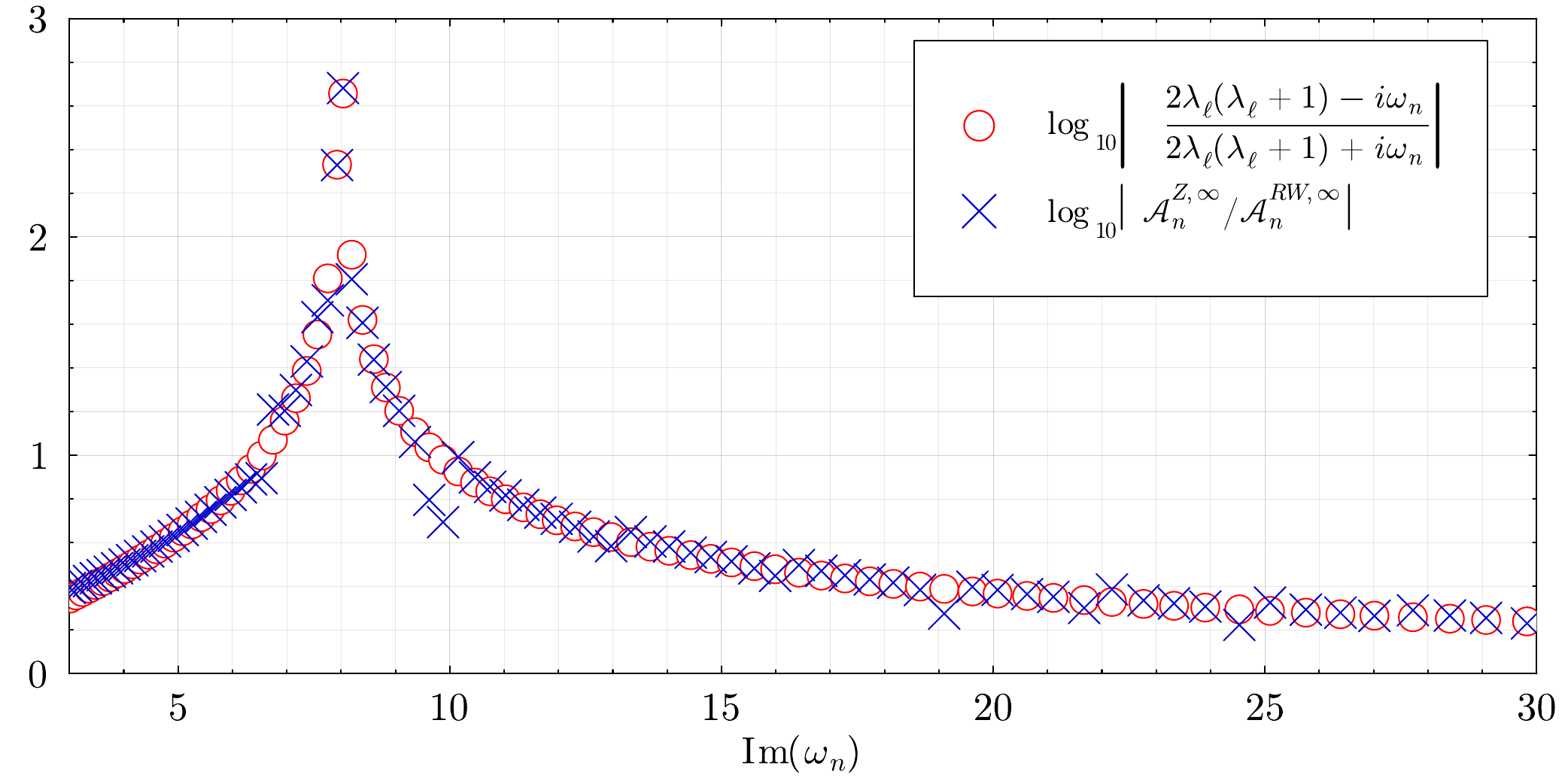}\label{CD:branch-cut}
    }
    \caption{Assessment of the ratio $|{\cal A}_n^{\infty,\mathrm{even}}(0)/{\cal A}_n^{\infty,\mathrm{odd}}(0)|$, for $\ell=2$. We proceed exactly as in Fig. \ref{f:CD_exact}, but
    considering the $\log_{10}$ of $|{\cal A}_n^{\infty,\mathrm{even}}(0)/{\cal A}_n^{\infty,\mathrm{odd}}(0)|$ 
    rather than that of $\left|B^{\mathrm{Kel,even}}_n/B^{\mathrm{Kel,odd}}_n\right|$. The agreement is very good for QNMs and indeed remarkable for branch-cut points, confirming again the consistency and robustness of the construction.    }
    \label{f:CD}
\end{figure}

Taking into account
the commented weak dependence on the initial-data  of the excitation pattern qualitative features  (see the point iv) above), and as a probe of the relative even-odd excitation with
only data at null infinity,
we directly compare the ratio of ${\cal A}_n^\infty(0)={\cal A}_n(x_{\scri^+})$'s in the even and odd-parity sectors, without normalising at the horizon and therefore subject to (a weak) dependence on initial data. Figure \ref{f:CD} shows 
$|{\cal A}_n^{\infty,\mathrm{even}}(0)/{\cal A}_n^{\infty,\mathrm{odd}}(0)|$, again for both QNMs and branch points
and comparing it with right-hand-side in Eq. (\ref{e:excitation_factors}). The agreement is again remarkable. 
Both Figs. \ref{f:CD_exact} and \ref{f:CD}
provide not only a check of the expected relative excitation 
of even and odd-parity modes, but a non-trivial and stringent test to the implemented hyperboloidal Keldysh resonant expansion.

Figures \ref{f:CD_exact} and \ref{f:CD} pose the following question: can the `Darboux' relation (\ref{e:excitation_factors}) between even and odd-parity excitations justify the observed `AS stagnation' effect in the even-parity case, i.e. the `anomalous' enhancement of AS frequencies as compared to nearby modes?
 
The answer is in the negative. Indeed, if we had a first-principles
justification for a `homogeneous' (monotonic without peaks) distribution in $n$ for the odd-parity amplitudes ${\cal A}_n^{\infty,\mathrm{odd}}(0)$, then the Darboux relation in Eq. (\ref{e:excitation_factors}) 
would indeed explain the enhancement of 
${\cal A}_n^{\infty,\mathrm{even}}(0)$ precisely at AS frequencies and, consequently, the observed 
`stagnation'. However, it could in principle have been the case 
that the even amplitudes ${\cal A}_n^{\infty,\mathrm{even}}(0)$
were `homogeneous' in $n$, so the odd amplitudes would have
been suppressed at AS frequencies, in accordance with Eq. (\ref{e:excitation_factors}). The fact is that the latter expression
only fixes the ratio of excitation between even-odd perturbations,
but says nothing about their respective intrinsic value.

At this point we recall the two points a) and b) highlighted in section \ref{s:AS_stagnation}, namely the relevant role played by both the branch cut and a (proper) AS QNM in the even-parity AS stagnation phenomenon. The AS QNM is absent in the odd-parity
case \cite{van2000analytic}, where no AS stagnation phenomenon is observed.
This points towards the simultaneous presence of the
branch cut and an AS QNM as a requirement for AS stagnation, suggesting an `interaction' between the AS QNM eigenvalue (point spectrum) with the branch cut (continuum spectrum) as underlying the mechanism behind the stagnation 
around the AS frequency. Such a spectral interaction would explain the absence of excitation enhancement in the
odd-parity case and it would also predict the absence of the AS stagnation phenomenon when a branch cut is not present. 
We will address this last point in the next section,
when looking to other spacetime asymptotics.
Such a (conjectured) spectral interaction description echoes the
``bound states in the continuum" (BIC) phenomenon in quantum mechanics \cite{von1993merkwurdige,hsu2016bound}
with a plausible metastable state accounting for the (first peak) `prompt-to-ringdown' phase.

\section{Brief remarks on other spacetime asymptotics and BH geometries}
\label{s:beyond_Schw}
A natural question concerns the presence of the `AS stagnation' phenomenon beyond Schwarzschild, either by changing the spacetime asymptotics to de Sitter and considering its null cosmological horizon (we do not consider Anti-de Sitter with its timelike null infinity), or 
by changing the BH geometry while staying in asymptotic flatness.
From the conjectured role of the interaction between 
an AS QNM eigenvalue and a continuum spectrum, one would expect the effect to be absent
in de Sitter asymptotics, due to the lack of branch cut.
Regarding other asymptotically flat geometries, with a focus on Reissner-Nordstr\"om and Kerr, we would also expect the effect to be absent in those classes of gravitational perturbations not presenting an AS QNM eigenvalue, while being present whenever  there indeed exists an AS QNM.

\begin{figure}[htp]
    \centering
    \subfloat[Schwarzschild-de Sitter Zerilli-like]{
      \includegraphics[clip,width=0.5\columnwidth]{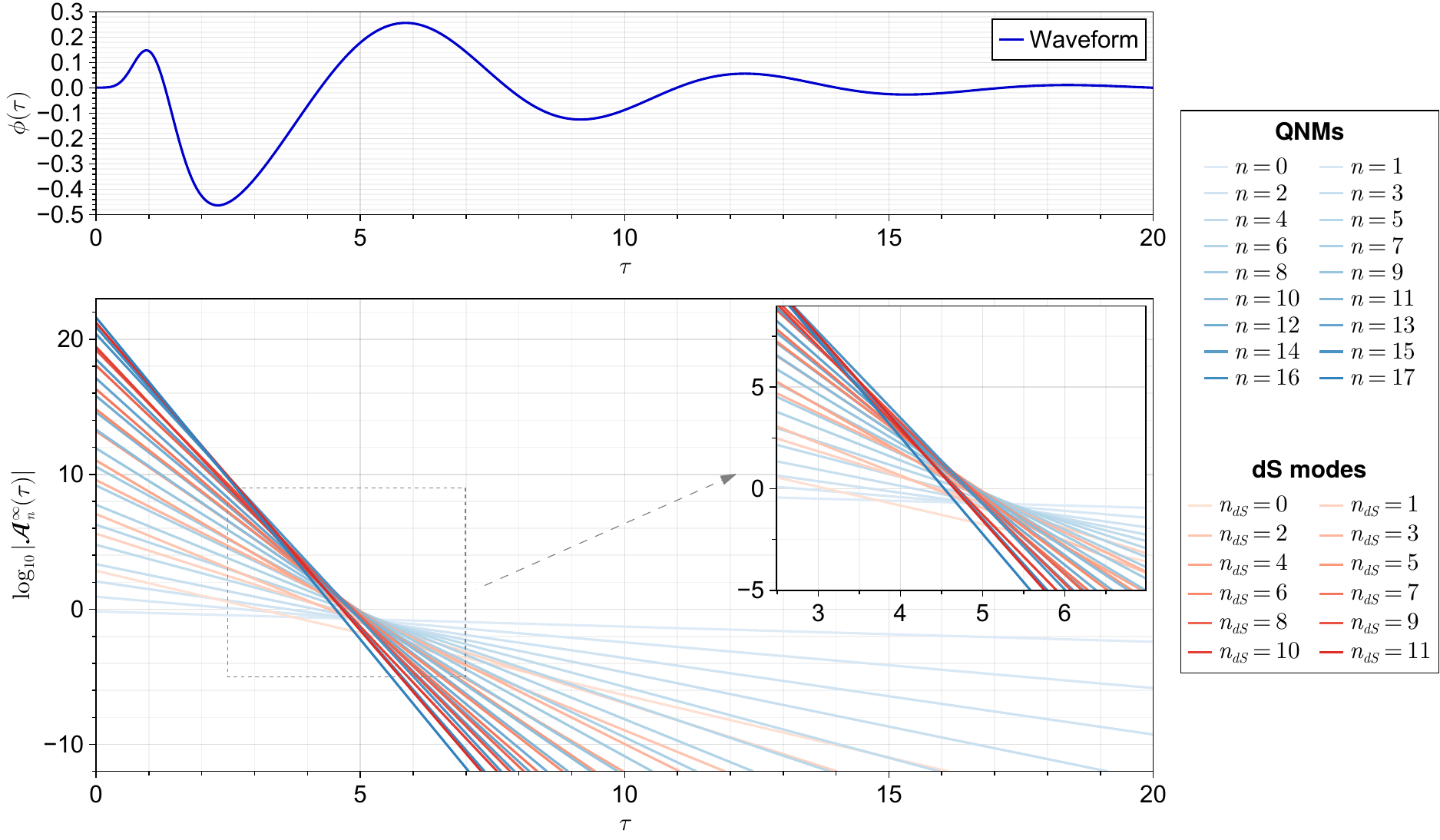}\label{panel:SdSZ:enveloping}
    }
    \subfloat[Schwarzschild-de Sitter Regge-Wheeler-like]{
      \includegraphics[clip,width=0.5\columnwidth]{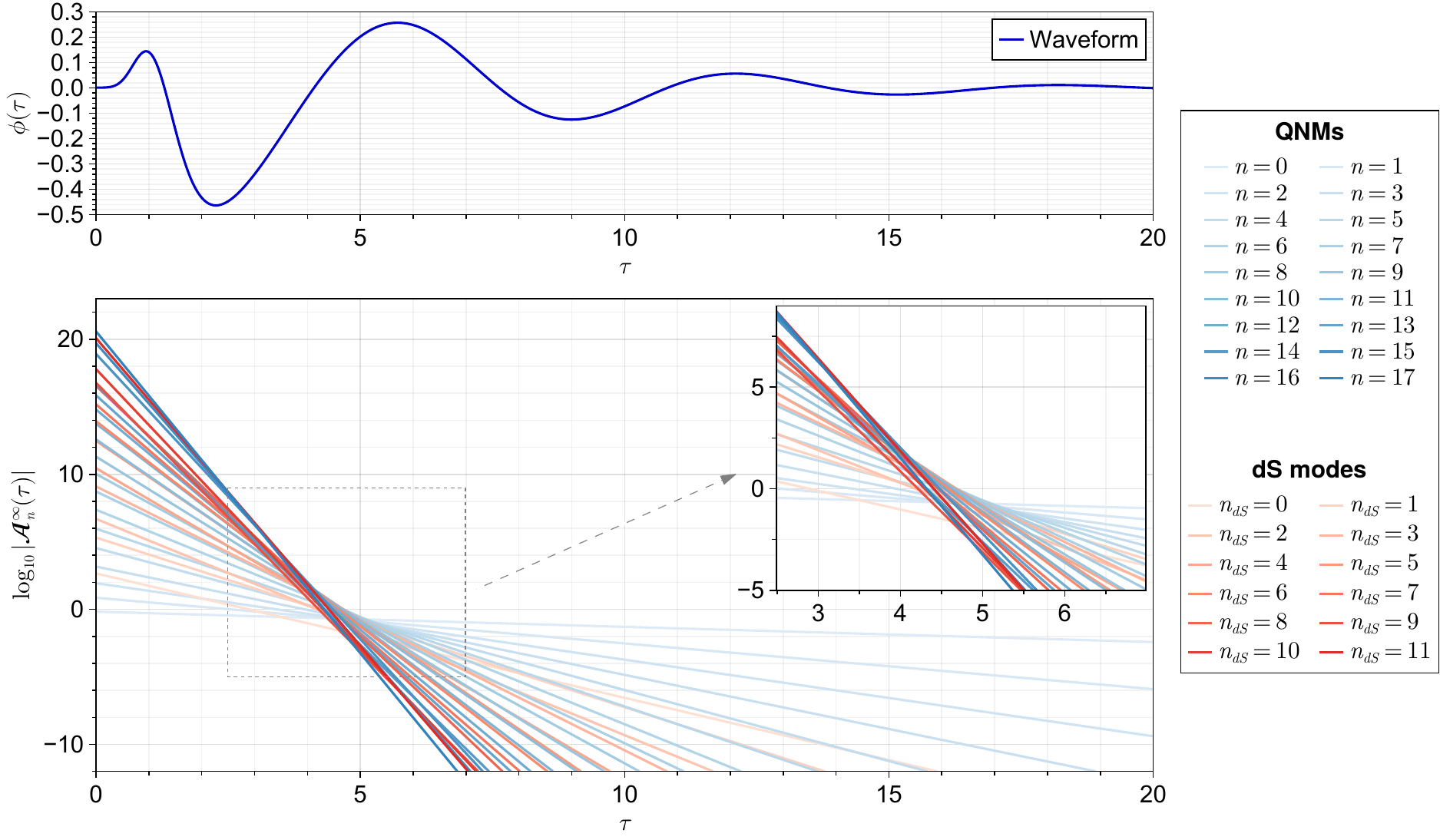}\label{panel:SdSRW:enveloping}
    }
    \caption{Evolution of $|{\cal A}^\infty_n(\tau)|$ in Schwarzschild-de Sitter ($\lambda=2r_{\cal H}, \Lambda M^2=0.07$). Absence of “secant-to-the-envelope” in both cases
    consistently with the absence of branch cut.}
    \label{f:dS}
\end{figure}

\begin{figure}[htp]
    \centering
    \subfloat[Reissner-Nordström Zerilli-like]{   \includegraphics[clip,width=0.5\columnwidth]{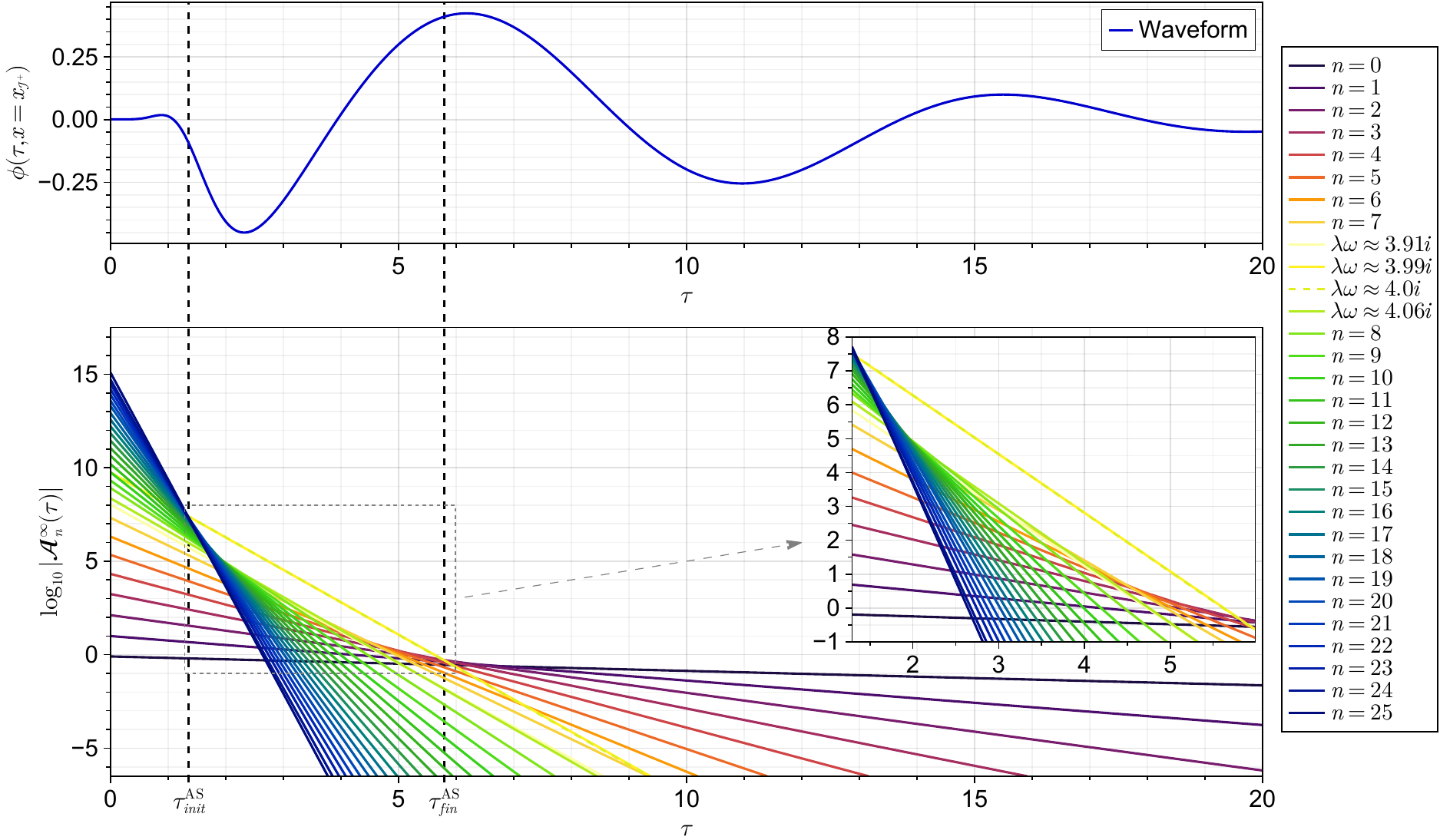}\label{panel:RNZ:enveloping}
    }
    \subfloat[Reissner-Nordström Regge-Wheeler-like]{
      \includegraphics[clip,width=0.5\columnwidth]{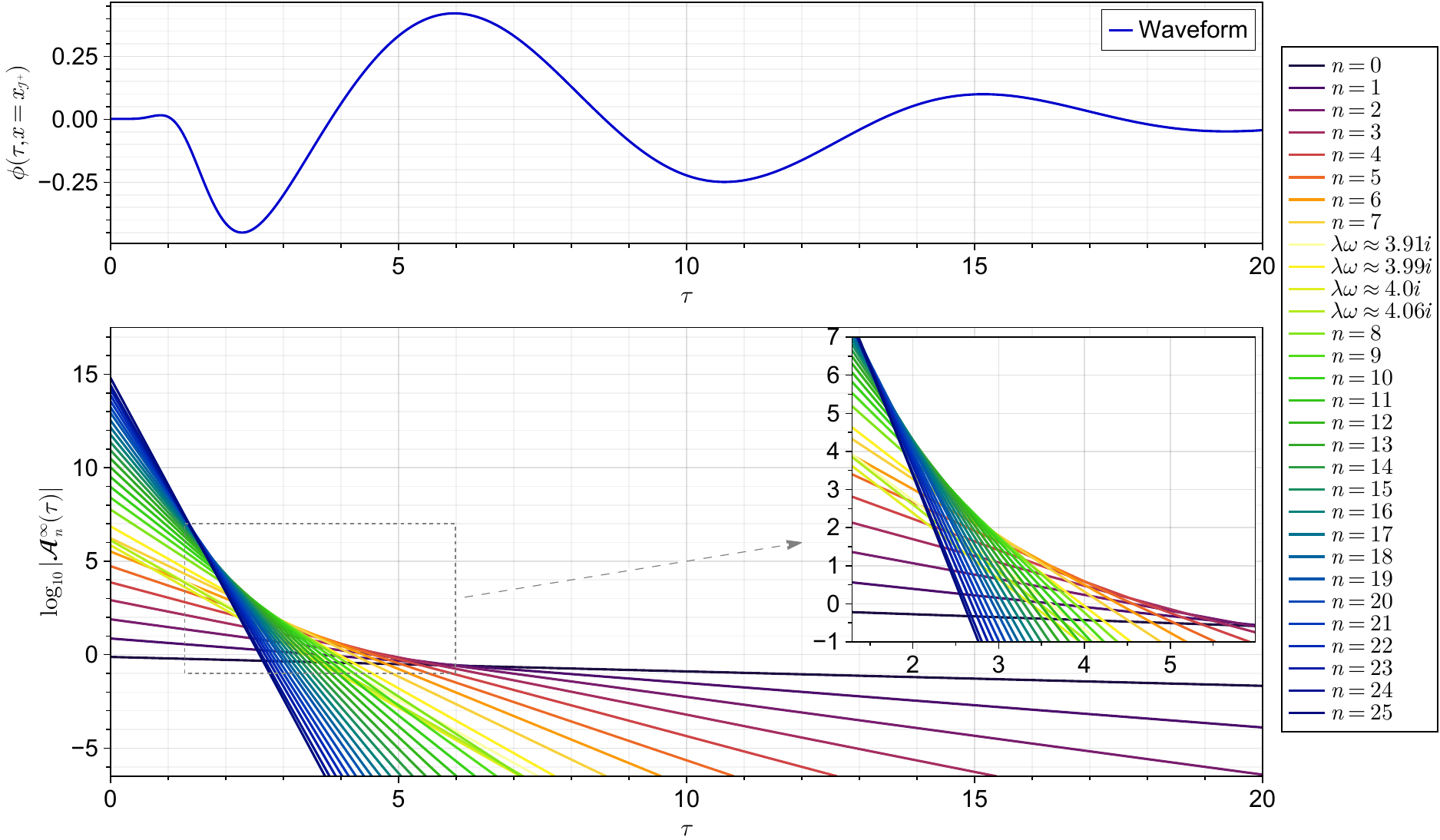}\label{panel:RNRW:enveloping}
    }
    \caption{Evolution of $|{\cal A}^\infty_n(\tau)|$, Reissner-Nordström ($\lambda =r_{\cal H}, Q/M=0.01$). Presence of “secant-to-the-envelope” for the (Zerilli-like) potential  $V_2^+$, 
    absence for (Regge-Wheeler-like) potential $V_2^-$, consistently with the respective presence/absence
    of an AS eigenvalue.}
    \label{f:RN}
\end{figure}

\begin{figure}[htp]
    \centering
    \subfloat[$s=-2$]{
      \includegraphics[clip,width=0.5\columnwidth]{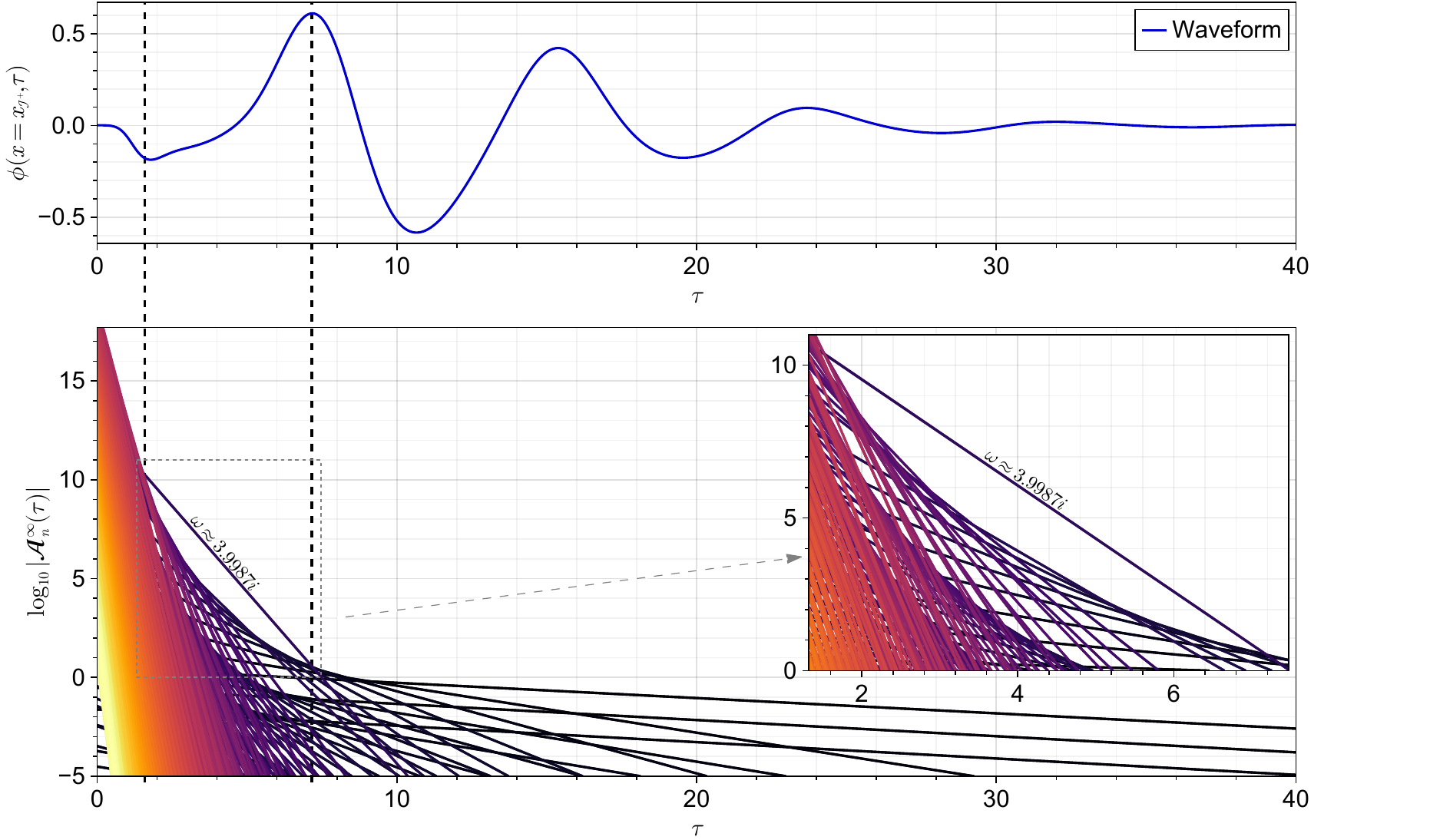}\label{panel:Kerr-:enveloping}
    }
    \subfloat[$s=2$]{
      \includegraphics[clip,width=0.5\columnwidth]{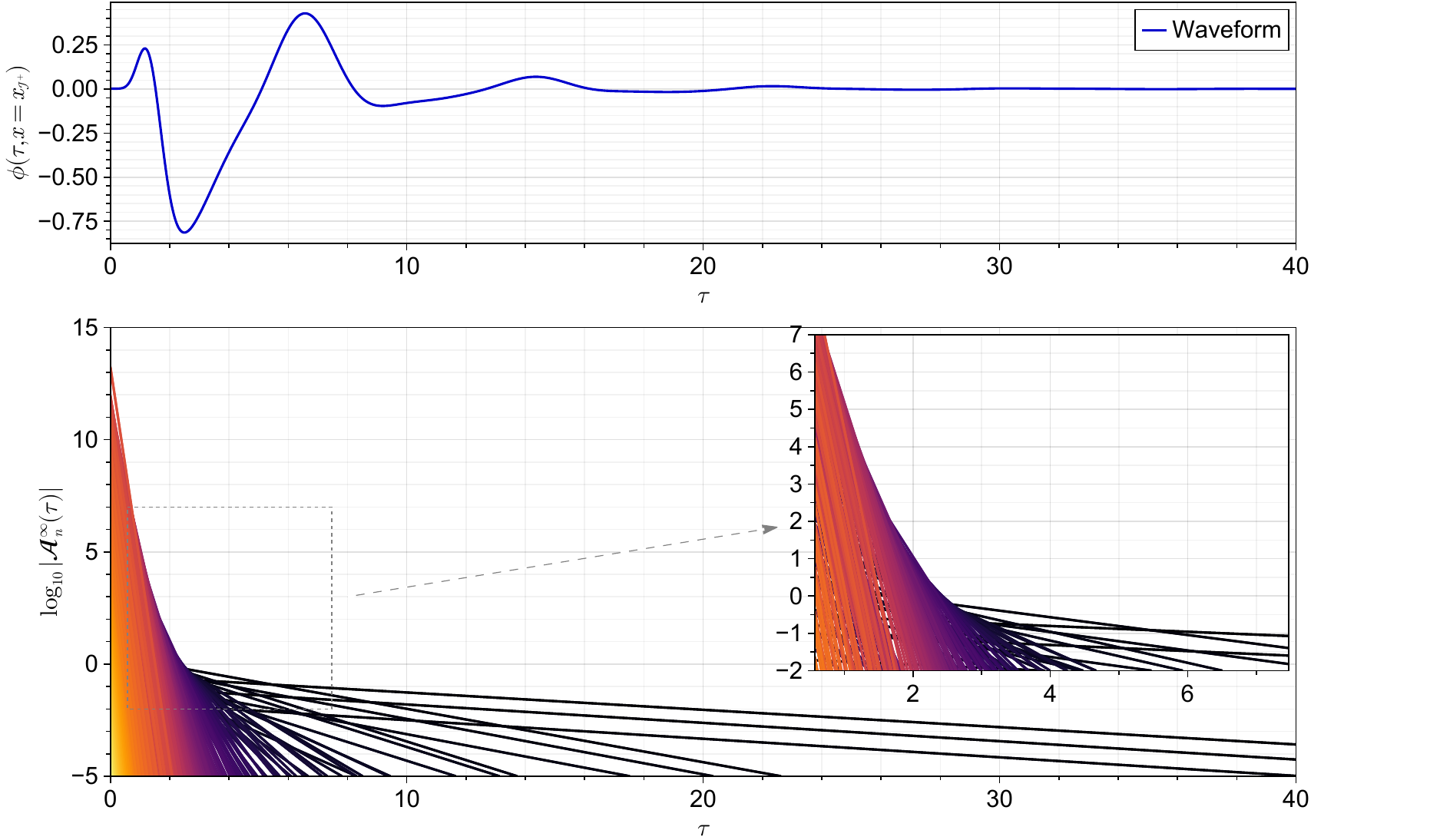}\label{panel:Kerr+:enveloping}
    }
\caption{Evolution of $|{\cal A}^\infty_n(\tau)|$ in Kerr ($\lambda=r_{\cal H},a/M=0.01$, for angular parameters $m=0$ and $\cos\theta=0$). Presence of the “secant-to-the-envelope” for the $s=-2$ potential and absence for the $s=2$ potential, consistently with the respective presence/absence of an AS QNM eigenvalue (modes are not labelled here, for the sake of preserving the visual clarity).}
\label{f:Kerr}
\end{figure}

We report here on the numerical results 
(see the supplementary material for details):
\begin{itemize}
    \item[1.] Schwarzschild-de Sitter (lengthscale $\lambda=2r_{\cal H}$):
    for a generic choice of cosmological constant $\Lambda$ we find an ``envelope" without a
        secant, that is, no `AS stagnation' in both Zerilli-like
        and Regge-Wheeler cases, consistently with the absence of continuum spectrum. This is illustrated in Fig. \ref{f:dS} for $\Lambda M^2=0.07$. 

However there exist special choices of $\Lambda$ for which the AS stagnation phenomenon occurs. In these special cases, the AS eigenfunction is more regular than in the generic case.
        This relation between regularity and enhanced excitation
        must be further explored.
        
    \item[2.] Other asymptotically flat BH geometries:
    \begin{itemize}
        \item[a)] Reissner-Nordstr\"om ($\lambda =r_{\cal H}$): presence of a ``secant-to-the-envelope" for the $V_2^+$ potential in Eq. (189), page 234 in \cite{Chandrasekhar:579245}, reducing to 
        Zerilli for $Q=0$. Absence for the
        odd-parity potential $V_2^-$ .
        This is illustrated in Fig.~\ref{f:RN} for $Q/M=0.01$.
         \item[b)] Kerr ($\lambda =r_{\cal H}$): presence of a ``secant-to-the-envelope" for gravitational $s=-2$ perturbations and absence in the case $s=2$ (see \cite{PanossoMacedo:2020biw,cai_pseudospectrum_2025, assaad_quasinormal_2025} for the hyperboloidal slicing and the numerical implementation of the QNM eigenvalue problem; see also \cite{van2000analytic,cook_modes_2016} for a discussion of the presence/absence of AS QNMs
         for the cases $s=\mp2$, and their relation to TTMs). This behaviour is illustrated in Fig.~\ref{f:Kerr} 
         for $a/M=0.01$.
    \end{itemize}
    
\end{itemize}

The consistency of these results beyond the Schwarzschild BH, namely the presence of a ``secant-to-the-envelope" (and therefore an `AS stagnation' phenomenon) whenever both an AS QNM eigenvalue and a branch cut are simultaneously realised by the BH geometry, while otherwise being absent in the generic case (though special BH de Sitter cases with increased regularity AS eigenfunctions do depart from this rule), provides  support to our intuition about  a mechanism involving the interaction between an AS QNM eigenvalue and a continuum (branch cut) spectrum, to account for the (anomalous)
enhanced activation of modes near the AS frequency. It would be interesting to assess the possible connection with phenomena resulting from the interaction of the AS QNM with modes in the `unconventional' sheet of the Riemann surface~\cite{leung2003unconventional,kubota2026pole}.
A detailed study of AS eigenmode regularity is indeed required.

\section{Conclusions and perspectives}
We have reported on the monitoring of the evolution in time  of the activation of QNMs (and the branch cut) in Vishveshwara-like linear scattered waveforms in Schwarzschild and beyond,
with different spacetime asymptotics and geometries (including
Reissner-Nordstr\"om and Kerr), by implementing the Keldysh 
spectral decomposition in \cite{besson_quasi-normal_2025}.
We have found that, for those cases presenting both an AS QNM (eigenvalue, point spectrum) and a branch cut (continuum spectrum), the modes close to the AS frequency are anomalously enhanced  in a universal manner not depending on the
initial conditions of the wavepacket. Such modes (QNM and branch cut) close to the AS frequency dominate in a dynamical phase after the initial prompt response that correlates in time with specific structures (peaks) in the waveform. In such cases, the end of the AS dominated phase provides the onset of a dynamical phase that presents the expected features of the
``ringdown" stage in non-linear BH evolutions.

We conjecture that the mechanism underlying the enhancement of modes close to the AS frequency involves the interaction between the AS QNM eigenvalue and the continuum spectrum, giving rise to a metastable state (in the spirit of a (quasi-)BIC) that controls an intermediate ``prompt-to-ringdown" dynamical phase.
 It is also tantalizing
to consider the possibility of a non-linear signature of this phenomenon, in particular aiming at formulating a sound characterisation of the ``ringdown onset" time in proper full non-linear BH dynamics.

As perspectives, pushing Vishveshwara's methodology
``on the BH trail", we consider:
\begin{itemize}
    \item[i)] To assess the conjecture on the BIC-like
    mechanism underlying the AS dynamical phase.
    This study comes together with a detailed analysis of the regularity of AS eigenmodes.
    
    \item[ii)] To study algebraically special perturbations of Schwarzschild in the Robinson-Trautman family, as the basis 
    for a second-order perturbation exploration of this AS phenomenon.
  
    \item[iii)] To explore the possibility of using  insights from this linear `AS stagnation' phenomenon in order to characterise the `ringdown onset' in non-linear head-on binary BH mergers.

\end{itemize}

\bigskip

\section*{Acknowledgments}
We would like to thank Michele Lenzi, Carlos F. Sopuerta and
Corentin Vitel for discussions in connection with Darboux transformations, KdV integrability and the `wave-mean flow' approach. We thank Jibril Ben Achour and
Gustavo Dotti for discussions on algebraically special perturbations of Schwarzschild. We thank  Sachindeo Vaidya for
key discussions and insights in connection with the BIC mechanism.
We also thank 
Piotr Bizo\'n, Alberto Garc\'\i a Mart\'\i n-Caro,  Daniel Pook-Kolb and Claude Warnick for discussions involving solitons and/or BH non-linear stability.
JLJ would like to warmly thank Saraswati and Smitha Vishveshwara for their kind sharing of thoughts and memories, as well as Juan Mar\'\i a Aguirregabiria and Bala Iyer for sharing their remembrances on Vishu's work. JLJ would also like to thank the organisers and colleagues at the conference ``Symphony of Spacetime
(Celebrating the Discovery of Black Hole Quasinormal Modes by C. V. Vishveshwara)", at the
Raman Research Institute (Bengaluru, India), 10-12 June 2026.
 We acknowledge support
from the EIPHI Graduate school (contract ``ANR-17-EURE-0002”), and in particular through the EIPHI-2025 project ``Resonan-SIES”. We also thank
the project QuanTEdu-France (22-CMAS-0001), the ``Investissements d'Avenir'' program through project ISITE-BFC (ANR-15-IDEX-03), the ANR
``Quantum Fields interacting with Geometry'' (QFG) project
(ANR-20-CE40-0018-02), and the Spanish FIS2017-86497-C2-1 project (with FEDER contribution). We finally thank the Erwin Schr\"odinger International Institute for Mathematics and
Physics (ESI) where part of this work has been developed during the Workshop on
“Hyperboloidal Foliations and their Application”.

\section*{Supplementary material}
Further supporting material can be found in the presentation \url{https://bessonjeremy.github.io/ModalWeightKeldysh/walkthrough.html}

\bigskip 

\bibliographystyle{reporthackunsrt}
\bibliography{Biblio}

@string{prl="Phys. Rev. Lett."}

@article{Abbott:2016blz,
  author        = {Abbott, B. P. and others},
  collaboration = {LIGO Scientific, Virgo},
  title         = {{Observation of Gravitational Waves from a Binary Black Hole Merger}},
  eprint        = {1602.03837},
  archiveprefix = {arXiv},
  primaryclass  = {gr-qc},
  reportnumber  = {LIGO-P150914},
  doi           = {10.1103/PhysRevLett.116.061102},
  journal       = {Phys. Rev. Lett.},
  volume        = {116},
  number        = {6},
  pages         = {061102},
  year          = {2016}
}

@article{Ansorg:2016ztf,
  author        = {Ansorg, Marcus and Panosso Macedo, Rodrigo},
  title         = {{Spectral decomposition of black-hole perturbations on
                   hyperboloidal slices}},
  journal       = {Phys. Rev.},
  volume        = {D93},
  year          = {2016},
  number        = {12},
  pages         = {124016},
  doi           = {10.1103/PhysRevD.93.124016},
  eprint        = {1604.02261},
  archiveprefix = {arXiv},
  primaryclass  = {gr-qc},
  slaccitation  = {%%CITATION = ARXIV:1604.02261;%%}
}

@article{assaad_quasinormal_2025,
  title        = {Quasinormal modes in Kerr spacetime as a 2D eigenvalue problem},
  volume       = {42},
  issn         = {0264-9381, 1361-6382},
  url          = {https://iopscience.iop.org/article/10.1088/1361-6382/ae24da},
  doi          = {10.1088/1361-6382/ae24da},
  pages        = {245008},
  number       = {24},
  journaltitle = {Classical and Quantum Gravity},
  shortjournal = {Class. Quantum Grav.},
  author       = {Assaad, Jamil and Panosso Macedo, Rodrigo},
  urldate      = {2026-08-01},
  date         = {2025-12-19}
}

@article{Berti:2009kk,
  author        = {Berti, Emanuele and Cardoso, Vitor and Starinets, Andrei O.},
  archiveprefix = {arXiv},
  doi           = {10.1088/0264-9381/26/16/163001},
  eprint        = {0905.2975},
  journal       = {Class.\ Quant.\ Grav.},
  pages         = {163001},
  primaryclass  = {gr-qc},
  title         = {{Quasinormal modes of black holes and black branes}},
  volume        = {26},
  year          = {2009}
}

@article{besson_quasi-normal_2025,
  title        = {Quasi-normal mode expansions of black hole perturbations: a hyperboloidal Keldysh’s approach},
  volume       = {57},
  issn         = {1572-9532},
  url          = {https://doi.org/10.1007/s10714-025-03438-6},
  doi          = {10.1007/s10714-025-03438-6},
  shorttitle   = {Quasi-normal mode expansions of black hole perturbations},
  aabstract    = {},
  pages        = {110},
  number       = {7},
  journaltitle = {General Relativity and Gravitation},
  shortjournal = {Gen Relativ Gravit},
  author       = {Besson, Jérémy and Jaramillo, José Luis},
  urldate      = {2025-09-01},
  date         = {2025-07-14},
  langid       = {english}
}

@article{cai_pseudospectrum_2025,
  title        = {Pseudospectrum for the Kerr black hole with spin s=0 case},
  volume       = {111},
  url          = {https://link.aps.org/doi/10.1103/PhysRevD.111.084011},
  doi          = {10.1103/PhysRevD.111.084011},
  pages        = {084011},
  number       = {8},
  journaltitle = {Physical Review D},
  shortjournal = {Phys. Rev. D},
  publisher    = {American Physical Society},
  author       = {Cai, Rong-Gen and Cao, Li-Ming and Chen, Jia-Ning and Guo, Zong-Kuan and Wu, Liang-Bi and Zhou, Yu-Sen},
  urldate      = {2025-09-01},
  date         = {2025-04-04}
}

@article{ChaDet75,
  issn      = {00804630},
  url       = {http://www.jstor.org/stable/78902},
  author    = {S. Chandrasekhar and S. Detweiler},
  journal   = {Proceedings of the Royal Society of London. Series A, Mathematical and Physical Sciences},
  number    = {1639},
  pages     = {441--452},
  publisher = {The Royal Society},
  title     = {The Quasi-Normal Modes of the Schwarzschild Black Hole},
  volume    = {344},
  year      = {1975}
}

@book{Chandrasekhar:579245,
  author    = {Chandrasekhar, S},
  title     = {{The mathematical theory of black holes}},
  publisher = {Oxford Univ. Press},
  address   = {Oxford},
  series    = {Oxford classic texts in the physical sciences},
  year      = {2002},
  url       = {https://cds.cern.ch/record/579245}
}

@article{cook_modes_2016,
  title        = {Modes of the Kerr geometry with purely imaginary frequencies},
  volume       = {94},
  rights       = {http://link.aps.org/licenses/aps-default-license},
  issn         = {2470-0010, 2470-0029},
  url          = {https://link.aps.org/doi/10.1103/PhysRevD.94.104074},
  doi          = {10.1103/PhysRevD.94.104074},
  pages        = {104074},
  number       = {10},
  journaltitle = {Physical Review D},
  shortjournal = {Phys. Rev. D},
  author       = {Cook, Gregory B. and Zalutskiy, Maxim},
  urldate      = {2026-07-30},
  date         = {2016-11-30}
}

@book{dyatlov2019mathematical,
  title     = {Mathematical Theory of Scattering Resonances},
  author    = {Dyatlov, S. and Zworski, M.},
  isbn      = {9781470443665},
  lccn      = {2019006281},
  series    = {Graduate Studies in Mathematics},
  url       = {https://books.google.fr/books?id=atCuDwAAQBAJ},
  year      = {2019},
  publisher = {American Mathematical Society}
}

@article{hintz_quasinormal_2021,
  title        = {Quasinormal modes and dual resonant states on de Sitter space},
  volume       = {104},
  issn         = {2470-0010, 2470-0029},
  url          = {https://link.aps.org/doi/10.1103/PhysRevD.104.064037},
  doi          = {10.1103/PhysRevD.104.064037},
  pages        = {064037},
  number       = {6},
  journaltitle = {Physical Review D},
  shortjournal = {Phys. Rev. D},
  author       = {Hintz, Peter and Xie, {YuQing}},
  urldate      = {2026-04-21},
  date         = {2021-09-14},
  langid       = {english}
}

@article{hintz2026nonlinear,
  title   = {Nonlinear stability of subextremal Kerr black holes},
  author  = {Hintz, Peter},
  journal = {arXiv preprint arXiv:2606.28253},
  year    = {2026}
}

@article{hsu2016bound,
  title     = {Bound states in the continuum},
  author    = {Hsu, Chia Wei and Zhen, Bo and Stone, A Douglas and Joannopoulos, John D and Solja{\v{c}}i{\'c}, Marin},
  journal   = {Nature Reviews Materials},
  volume    = {1},
  number    = {9},
  pages     = {16048},
  year      = {2016},
  publisher = {Nature Publishing Group}
}

@article{Jaramillo:2020tuu,
  author        = {Jaramillo, Jos\'e Luis and Panosso Macedo, Rodrigo and Al Sheikh, Lamis},
  title         = {{Pseudospectrum and Black Hole Quasinormal Mode Instability}},
  eprint        = {2004.06434},
  archiveprefix = {arXiv},
  primaryclass  = {gr-qc},
  doi           = {10.1103/PhysRevX.11.031003},
  journal       = {Phys. Rev. X},
  volume        = {11},
  number        = {3},
  pages         = {031003},
  year          = {2021}
}

@article{Jaramillo:2023day,
  author        = {Jaramillo, Jos\'e Luis and Krishnan, Badri and Sopuerta, Carlos F.},
  title         = {{Universality in binary black hole dynamics: An integrability conjecture}},
  eprint        = {2305.08554},
  archiveprefix = {arXiv},
  primaryclass  = {gr-qc},
  doi           = {10.1142/S0218271823420051},
  journal       = {Int. J. Mod. Phys. D},
  volume        = {32},
  number        = {14},
  pages         = {2342005},
  year          = {2023}
}

@book{LaxPhi89,
  optkey      = {LaxPhi89},
  author      = {Lax, P. D. and Phillips, R. S},
  title       = {Scattering theory},
  publisher   = {Academic Press},
  year        = {1989},
  optcrossref = {},
  opteditor   = {},
  volume      = {26},
  optnumber   = {},
  series      = {Pure and Applied Mathematics},
  address     = {Boston},
  edition     = {Second edition},
  optmonth    = {},
  optnote     = {},
  optannote   = {}
}

@article{Leaver85,
  key     = {Leaver85},
  author  = {E.W. Leaver},
  title   = {An analytic representation for the quasi-normal modes of
             {K}err black holes},
  journal = {Proc. R. Soc. London, Ser. A},
  volume  = {402},
  year    = {1985},
  pages   = {285-298}
}

@article{MacZen25,
  author   = {Panosso Macedo, Rodrigo  and Zenginoğlu, Anıl },
  title    = {Hyperboloidal approach to quasinormal modes},
  journal  = {Frontiers in Physics},
  volume   = {Volume 12 - 2024},
  year     = {2025},
  url      = {https://www.frontiersin.org/journals/physics/articles/10.3389/fphy.2024.1497601},
  doi      = {10.3389/fphy.2024.1497601},
  issn     = {2296-424X}
}

@article{PanossoMacedo:2018hab,
  author        = {Panosso Macedo, Rodrigo and Jaramillo, Jos\'e Luis and
                   Ansorg, Marcus},
  title         = {{Hyperboloidal slicing approach to quasi-normal mode
                   expansions: the Reissner-Nordstr\"om case}},
  journal       = {Phys. Rev.},
  volume        = {D98},
  year          = {2018},
  number        = {12},
  pages         = {124005},
  doi           = {10.1103/PhysRevD.98.124005},
  eprint        = {1809.02837},
  archiveprefix = {arXiv},
  primaryclass  = {gr-qc},
  slaccitation  = {%%CITATION = ARXIV:1809.02837;%%}
}

@article{PanossoMacedo:2020biw,
  author       = {Panosso Macedo, Rodrigo},
  title        = {{Hyperboloidal framework for the Kerr spacetime}},
  journal      = {Class. Quant. Grav.},
  volume       = {37},
  year         = {2020},
  number       = {6},
  pages        = {065019},
  doi          = {10.1088/1361-6382/ab6e3e},
  slaccitation = {%%CITATION = CQGRD,37,065019;%%}
}

@article{Pre05,
  author    = {F. Pretorius},
  title     = { Evolution of Binary Black-Hole Spacetimes},
  journal   = prl,
  year      = {2005},
  optkey    = {},
  volume    = {95},
  optnumber = {},
  pages     = {121101},
  optmonth  = {},
  optnote   = {},
  optannote = {}
}

@article{Szeftel:2026qzr,
  author  = {Szeftel, J{\'e}r{\'e}mie},
  title   = {{Kerr stability in the full subextremal range}},
  journal = {arXiv preprint arXiv:2609.40068},
  year    = {2026}
}

@article{van2000analytic,
  title     = {Analytic treatment of black-hole gravitational waves at the algebraically special frequency},
  author    = {van den Brink, Alec Maassen},
  journal   = {Physical Review D},
  volume    = {62},
  number    = {6},
  pages     = {064009},
  year      = {2000},
  publisher = {APS}
}

@article{Vishveshwara_CurrentScience:1996jgz,
  author  = {Vishveshwara, C. V.},
  editor  = {Date, G. and Iyer, Bala R.},
  title   = {{On the black hole trail...: A personal journey}},
  journal = {Curr. Sci.},
  volume  = {71},
  number  = {11},
  pages   = {824--830},
  year    = {1996}
}

@phdthesis{Vishveshwara1968,
  author = {Vishveshwara, C. V.},
  title  = {The Stability of the {S}chwarzschild Metric},
  school = {University of Maryland, College Park},
  year   = {1968},
  type   = {{PhD} thesis}
}

@article{vishveshwara1970scattering,
  title     = {Scattering of gravitational radiation by a Schwarzschild black-hole},
  author    = {Vishveshwara, C.V.},
  journal   = {Nature},
  volume    = {227},
  number    = {5261},
  pages     = {936--938},
  year      = {1970},
  publisher = {Nature Publishing Group UK London}
}

@article{vishveshwara1970stability,
  title     = {Stability of the Schwarzschild metric},
  author    = {Vishveshwara, C.V.},
  journal   = {Physical Review D},
  volume    = {1},
  number    = {10},
  pages     = {2870},
  year      = {1970},
  publisher = {APS}
}

@incollection{von1993merkwurdige,
  title     = {{\"U}ber merkw{\"u}rdige diskrete Eigenwerte},
  author    = {von Neumann, John and Wigner, Eugene P},
  booktitle = {The Collected Works of Eugene Paul Wigner: Part A: The Scientific Papers},
  pages     = {291--293},
  year      = {1993},
  publisher = {Springer}
}

@article{Warnick:2013hba,
  author        = {Warnick, Claude M.},
  title         = {{On quasinormal modes of asymptotically anti-de Sitter
                   black holes}},
  journal       = {Commun. Math. Phys.},
  volume        = {333},
  year          = {2015},
  number        = {2},
  pages         = {959-1035},
  doi           = {10.1007/s00220-014-2171-1},
  eprint        = {1306.5760},
  archiveprefix = {arXiv},
  primaryclass  = {gr-qc},
  reportnumber  = {ALBERTA-THY-3-13},
  slaccitation  = {%%CITATION = ARXIV:1306.5760;%%}
}

@article{Zenginoglu:2011jz,
  author        = {Zenginoglu, Anil},
  title         = {{A Geometric framework for black hole perturbations}},
  journal       = {Phys. Rev.},
  volume        = {D83},
  year          = {2011},
  pages         = {127502},
  doi           = {10.1103/PhysRevD.83.127502},
  eprint        = {1102.2451},
  archiveprefix = {arXiv},
  primaryclass  = {gr-qc},
  slaccitation  = {%%CITATION = ARXIV:1102.2451;%%}
}

@article{ZhangBertiCardoso13,
  title     = {Quasinormal ringing of Kerr black holes. II. Excitation by particles falling radially with arbitrary energy},
  author    = {Zhang, Zhongyang and Berti, Emanuele and Cardoso, Vitor},
  journal   = {Phys. Rev. D},
  volume    = {88},
  issue     = {4},
  pages     = {044018},
  numpages  = {15},
  year      = {2013},
  month     = {Aug},
  publisher = {American Physical Society},
  doi       = {10.1103/PhysRevD.88.044018},
  url       = {https://link.aps.org/doi/10.1103/PhysRevD.88.044018}
}

@article{leung2003unconventional,
  title={Unconventional gravitational excitation of a Schwarzschild black hole},
  author={Leung, PT and van den Brink, Alec Maassen and Mak, KW and Young, K},
  journal={Classical and Quantum Gravity},
  volume={20},
  number={16},
  pages={L217--L223},
  year={2003}
}

@article{kubota2026pole,
  title={Pole Skipping, Avoided Crossing, and Resonant Excitation in Kerr Quasinormal Modes near Algebraically Special Frequencies},
  author={Kubota, Kei-ichiro and Motohashi, Hayato},
  journal={arXiv preprint arXiv:2605.17840},
  year={2026}
}

\end{document}